\documentclass[twocolumn,superscriptaddress,floatfix,aps,prd,nofootinbib,showpacs]{revtex4}

\usepackage{graphicx}

\begin{document}

\title{Black hole-neutron star binaries in general relativity:
  effects of neutron star spin}

\author{Keisuke Taniguchi}
\affiliation{Department of Physics, University of Illinois at
  Urbana-Champaign, Urbana, Illinois 61801, USA}
\author{Thomas W. Baumgarte}
\altaffiliation{Fellow of the J.~S.~Guggenheim Memorial Foundation}
\altaffiliation{Also at: Department of Physics, University of Illinois at
  Urbana-Champaign, Urbana, Illinois 61801, USA}
\affiliation{Department of Physics and Astronomy, Bowdoin College,
  Brunswick, Maine 04011, USA}
\author{Joshua A. Faber}
\altaffiliation{National Science Foundation (NSF) Astronomy and
  Astrophysics Postdoctoral Fellow}
\affiliation{Department of Physics, University of Illinois at
  Urbana-Champaign, Urbana, Illinois 61801, USA}
\author{Stuart L. Shapiro}
\altaffiliation{Also at: Department of Astronomy and NCSA, University of
  Illinois at Urbana-Champaign, Urbana, Illinois 61801, USA}
\affiliation{Department of Physics, University of Illinois at
  Urbana-Champaign, Urbana, Illinois 61801, USA}

\date{October 24, 2007}

\begin{abstract}

We present new sequences of general relativistic, quasiequilibrium 
black hole-neutron star binaries.
We solve for the gravitational field in the conformal thin-sandwich
decomposition of Einstein's field equations, coupled to the equations of
relativistic hydrostatic equilibrium for a perfect fluid.
We account for the black hole by solving these equations
in the background metric of a Schwarzschild black hole
whose mass is much larger than that of the neutron star.
The background metric is treated in Kerr-Schild as well as isotropic
coordinates.
For the neutron star, we assume a polytropic equation of state with
adiabatic index $\Gamma=2$, and solve for both irrotational and
corotational configurations.
By comparing the results of irrotational and synchronized configurations
with the same background metric, we conclude that the effect of the
rotation on the location of tidal break-up is only on the order
of a few percent.
The different choices in the background also lead to differences of order
a few percent, which may be an indication of the level to which these
configurations approximate quasiequilibrium.

\end{abstract}


\maketitle

\section{Introduction}

Black hole-neutron star (hereafter BHNS) binary systems are, together with
other compact binaries, among the most promising sources of gravitational
waves for detection by ground-based interferometers such as
LIGO \cite{LIGO}, GEO600 \cite{GEO600}, TAMA300 \cite{TAMA300},
and VIRGO \cite{VIRGO}, or for the planned space-based mission
LISA \cite{LISA}.
Since theoretical predictions for the waveforms from such compact binaries
are needed both for the identification and interpretation of any
astrophysical signals, significant effort has gone into the theoretical
modeling of these binaries and their inspiral
(see \cite{Blanc02,BaumS03} for recent reviews).

The orbital separation of a BHNS binary decreases as energy
and angular momentum are dissipated by the emission of gravitational
radiation, which also has the effect of circularizing the orbit,
until the system eventually coalesces.
This coalescence can take two qualitatively different forms.
If the neutron star reaches the innermost
stable circular orbit (hereafter ISCO) while still stable against tidal
disruption, it will likely plunge and fall into the black hole promptly.
Alternately, if the neutron star is tidally disrupted outside the ISCO,
the final fate may be a black hole surrounded by a disk.
This model is a candidate central engine for  short-period gamma-ray bursts,
since the high efficiency of accretion onto a black hole can help to
explain the huge luminosities seen in these sources
\cite{JankaERF99,Rossw05,CMiller05}.
Additionally, we may obtain information about the equation of state
(hereafter EOS) of matter at nuclear densities through the detection of
a gravitational wave signal, because the characteristic frequency at the
tidal disruption separation is in the most sensitive range of the
ground-based detectors \cite{Valli00}.

The difference in the final fate is determined primarily by the mass ratio
of the black hole to the neutron star.
When the black hole mass $M_{\rm BH}$ is much larger than the neutron star
mass $M_{\rm NS}$, the tidal force from the black hole is not so large 
compared with the neutron star self-gravity to
disrupt the neutron star, even at the ISCO.
Indeed, the ISCO occurs at a radius proportional to the black hole
mass, whereas the Roche limit separation, which would mark the beginning
of mass transfer, scales like $M_{\rm BH}^{1/3}$ in the limit of an
extremely large mass ratio \cite{Paczy71}.
Thus, the larger the black hole mass, the further outward the ISCO
lies relative to the Roche limit separation.
On the other hand, a stellar mass black hole will have a tidal field
sufficiently large with respect to the neutron star self-gravity
to deform and disrupt the companion outside the ISCO.
The critical value of the mass ratio falls in the range
$M_{\rm BH}/M_{\rm NS} \sim 4$ for reasonable neutron star models
\cite{Shiba96}, but varies depending on assumptions about the neutron star
EOS (see \cite{FaberBSTR05} for a thorough discussion).

Until now, much effort has been devoted to the computation
of the BHNS binary systems.
Most dynamical simulations so far have been carried out in a Newtonian
framework \cite{JankaERF99,Mash75,CartL83,Marck83,LeeK99,Lee00,RossSW04},
but see \cite{KobaLPM04,FaberBSTR05} for approximate relativistic treatments.
Quasiequilibrium models of BHNS binaries have been constructed
adopting various different approximations.
Several authors (including \cite{LaiRS93}) have modeled BHNS binaries
as Newtonian ellipsoids around point masses, generalizing the classic
Roche model for incompressible stars \cite{Chand69}.
These ellipsoidal calculations have also been generalized to include
relativistic effects
\cite{Fish73,Shiba96,LaiW96,TanigN96,WiggL00}.
Recently, these models have been generalized to include higher order
deformations than ellipsoid by expanding the background black hole
metric to higher order than quadrupole \cite{IshiiSM05}.
Equilibrium models of Newtonian BHNS binaries also have been constructed
by solving the exact fluid equations numerically and again treating the
black hole as a point-mass \cite{UryuE99}.
With the possible exception of \cite{Miller01},
Baumgarte {\it et al.} \cite{BaumSS04} (hereafter BSS) so far provide
the only self-consistent relativistic treatment of BHNS binaries in
quasiequilibrium.

In BSS, BHNS binaries are constructed by solving the constraint
equations of general relativity, decomposed in the conformal
thin-sandwich formalism, together with the Euler equation for the
neutron star matter, which takes an algebraic form if the system can
be taken to be stationary in a corotating frame.
The equations are solved in the background of a Schwarzschild black hole,
which accounts for the neutron star's companion.

In this paper we generalize the findings of BSS in two ways.
As discussed above we allow of irrotational instead of corotational
fluid flow, which is more realistic astrophysically \cite{Kocha92,BildsC92}.
To do so we adopt the formalism for constructing irrotational stars
as developed in \cite{GourGTMB01}.
We also generalize the results of BSS by expressing the black hole
background in a different coordinate system.
In BSS, this background was expressed in Kerr-Schild coordinates, which we
compare here with a black hole background expressed in isotropic coordinates.
Since the two coordinate system represent different slicings of the
Schwarzschild space-time, the resulting initial data are physically distinct
solutions of the constraint equations.
Finally, in this paper we adopt a numerical code that is based on the
spectral method numerical libraries known as {\tt LORENE}
\cite{Lorene} (as opposed to the finite difference implementation of
BSS).  We maintain several of the other assumptions made by BSS,
including extreme mass ratio $M_{\rm BH} \gg M_{\rm NS}$ and polytropic
equations of state, and plan to relax these in the future.

The spectral techniques used here have previously been  employed to
compute quasiequilibrium sequences of binary neutron stars
\cite{GourGTMB01,TanigGB01,TanigG02a,TanigG02b,TanigG03,BejgGGHTZ05}.
There are several advantages in using spectral methods.
One is that it is easy to treat spherical coordinates,
which are suitable for solving figures like stars,
and to fit the coordinates to trace the stellar surface.
This fitting technique plays an important role in solving irrotational
configurations.
Another advantage is that it is possible to achieve more rapid convergence
compared to finite difference methods,
until configurations reach a separation very close to tidal break-up
and the appearance of discontinuities in physical quantities
(For more details on the spectral methods techniques we use, 
we refer the readers to \cite{BonaGM98,GranBGM01,GourGTMB01}.)
In our code, the set of equations we solve for the gravitational field
is essentially equivalent to that of BSS, while the hydrostatic equations
are the same as those found in \cite{GourGTMB01}.

The paper is organized as follows. In Section II, we briefly summarize
the formulation and explain the solution procedure.
The tests of the numerical code are shown in Section III,
and the results are presented in Section IV.
In Section V we discuss the effects of the choice of the black
hole background metric, and we summarize in Section VI.

Throughout the present paper, we adopt geometrical units, $G=c=1$,
where $G$ denotes the gravitational constant and $c$ the speed of light,
respectively.
Latin and Greek indices denote purely spatial and space-time
components, respectively.

\section{Formulation}

In this Section we briefly discuss the basic equations, as well as
their numerical implementation.
For a more detailed discussion, we refer the reader to Section III of BSS
for the gravitational fields (but point out some minor differences below)
and to Section II of \cite{GourGTMB01} for the hydrostatics.
For corotating sequences using the Kerr-Schild metric, our notation will
appear to be slightly different than that found in BSS, but the two sets of
equations are completely equivalent, merely expressed in a different
set of variables.

\subsection{Gravitational field equations}

The line element in 3+1 form is written as
\begin{eqnarray}
  ds^2 &=& g_{\mu \nu} dx^{\mu} dx^{\nu} \nonumber \\
  &=& -\alpha^2 dt^2 +\gamma_{ij} (dx^i +\beta^i dt) (dx^j +\beta^j dt),
\end{eqnarray}
where $\alpha$ is the lapse function, $\beta^i$ the shift vector,
$\gamma_{ij}$ the spatial metric, and $g_{\mu \nu}$ the space-time metric.
The Einstein equations then split into two constraint equations
-- the Hamiltonian and the momentum constraint --
and two evolution equations -- one for the spatial metric $\gamma_{ij}$
and one for the extrinsic curvature
\begin{equation}
  K_{ij} = - \frac{1}{2 \alpha} \left( \partial_t \gamma_{ij}
  + D_i \beta_j + D_j \beta_i \right), \label{eq:ext_curv}
\end{equation}
where $D_i$ denotes the covariant derivative associated with $\gamma_{ij}$.
Using the conformal decomposition
\begin{equation}
  \gamma_{ij} = \psi^4 \tilde{\gamma}_{ij},
\end{equation}
where $\psi$ is the conformal factor and $\tilde{\gamma}_{ij}$ the
background metric, the Hamiltonian constraint coupled with the trace part
of the evolution equation of $K^{ij}$ becomes
\begin{eqnarray}
  &&\tilde{D}^2 \sigma = 4\pi \psi^4 S
  +{3 \over 4} \psi^{-8} \tilde{A}_{ij} \tilde{A}^{ij}
  +{1 \over 4} \tilde{R} +{1 \over 2} \psi^4 K^2 \nonumber \\
  &&+ {\psi^4 \over \alpha} \beta^i \tilde{D}_i K
  -{1 \over 2} \tilde{\gamma}^{ij}
  \bigr[ (\tilde{D}_i \nu) (\tilde{D}_j \nu) +(\tilde{D}_i \sigma)
    (\tilde{D}_j \sigma) \bigr], \label{eq:lognp}
\end{eqnarray}
where we have defined
\begin{equation} 
  \sigma \equiv \ln (\alpha \psi^2); \label{eq:sigma}
\end{equation}
this quantity was denoted ``$\beta$'' in \cite{GourGTMB01}.
$\tilde{D}_i$, $\tilde{D}^2=\tilde{\gamma}^{ij} \tilde{D}_i
\tilde{D}_j$, $\tilde{R}_{ij}$, and $\tilde{R}$ are, respectively,
the covariant derivative, covariant Laplace operator, Ricci tensor,
and scalar curvature with respect to $\tilde{\gamma}_{ij}$.
We have also decomposed the extrinsic curvature into its trace and
traceless parts,
\begin{equation}
  K^{ij} = \psi^{-10} \tilde{A}^{ij} + \frac{1}{3} \gamma^{ij} K.
\end{equation} 
In the derivation of (\ref{eq:lognp}) we have assumed $\partial_t K =
0$, which results in
\begin{eqnarray}
 &&\tilde{D}^2 \nu = 4\pi \psi^4 (\rho +S)
  +\psi^{-8} \tilde{A}_{ij} \tilde{A}^{ij}
  +{1 \over 3} \psi^4 K^2 \nonumber \\
 &&+ {\psi^4 \over \alpha} \beta^i \tilde{D}_i K
  -\tilde{\gamma}^{ij} (\tilde{D}_i \nu) (\tilde{D}_j \sigma) \label{eq:logn}
\end{eqnarray}
for the quantity
\begin{equation}
  \nu \equiv \ln \alpha. \label{eq:nu}
\end{equation}
Finally, we assume that $\partial_t \tilde \gamma_{ij} = 0$ for
quasiequilibrium configuration, so that (\ref{eq:ext_curv}) yields
\begin{equation}
  \tilde A^{ij} = \frac{\psi^6}{2 \alpha} \left(
  \tilde D^i \beta^j + \tilde D^j \beta^i - \frac{2}{3} \tilde \gamma^{ij}
  \tilde D_k \beta^k \right). \label{eq:ext_curv_2}
\end{equation}
Inserting this relation into the momentum constraint yields
\begin{eqnarray}
  &&\tilde{D}^2 \beta^i + {1 \over 3} \tilde{D}^i \tilde{D}_j \beta^j =
  16 \pi \alpha \psi^4 j^i -\tilde{R}^i_j \beta^j \nonumber \\
  &&+ {4 \over 3} \alpha \tilde{D}^i K
  -{2\alpha \over \psi^6} \tilde{A}^{ij} \tilde{D}_j (3\sigma -4\nu).
  \label{eq:shift}
\end{eqnarray}
The matter quantities on the right-hand side of equations
(\ref{eq:lognp}), (\ref{eq:logn}) and (\ref{eq:shift}) are projections
of the stress-energy tensor
\begin{equation}
  T_{\mu \nu} =(\rho_0 +\rho_i +P) u_{\mu} u_{\nu} +P g_{\mu \nu},
\end{equation}
where $u_{\mu}$ is the fluid 4-velocity, $\rho_0$ the baryon rest-mass
density, $\rho_i$ the internal energy density, and $P$ the pressure.
We then define
\begin{eqnarray}
  \rho &\equiv& n_{\mu} n_{\nu} T^{\mu \nu}, \\
  j^i &\equiv& -\gamma^i_{\mu} n_{\nu} T^{\mu \nu}, \\
  S_{ij} &\equiv& \gamma_{i \mu} \gamma_{j \nu} T^{\mu \nu}, \\
  S &\equiv& \gamma^{ij} S_{ij},
\end{eqnarray}
where $n_{\mu}=(-\alpha,0,0,0)$ is the future-directed unit normal vector.

\subsection{Background black hole metric}

We account for the neutron star's black hole companion by choosing a
background solution that represents a Schwarzschild metric.
In BSS the Schwarzschild solution was expressed in Kerr-Schild
coordinates.
In this paper we analyze the effect of this choice by comparing with
the Schwarzschild background expressed in isotropic coordinates.
Since the two coordinate system represent two distinct slicings of the
Schwarzschild metric, we solve the constraint equations on different
spatial slices, and therefore have no reason to expect that the
resulting solutions to the constraint equations are physically identical.

\begin{table}[ht]
\caption{Lapse function $\alpha_{\rm BH}$, shift vector $\beta^i_{\rm BH}$,
conformal factor $\psi_{\rm BH}$, and the conformally related spatial
metric $\tilde{\gamma}_{ij}$ for the Schwarzschild metric in
Kerr-Schild (K-S) and isotropic (ISO) coordinates.
$M_{\rm BH}$ is the black hole mass,
$r_{\rm BH}=\sqrt{X^2+Y^2+Z^2}$ the distance from the black hole center,
and we define $H_{\rm BH} \equiv M_{\rm BH}/r_{\rm BH}$
and $l_i =l^i \equiv X^i/r_{\rm BH}$.
Note that $r_{\rm BH}$ in Kerr-Schild
coordinates indicates a different displacement than one measured 
in isotropic ones.}
\begin{center}
\begin{tabular}{c|cc} \hline\hline
                   &K-S&ISO \\ \hline
  $\alpha_{\rm BH}$&$(1+2H_{\rm BH})^{-1/2}$&
  $\displaystyle {1-H_{\rm BH}/2 \over 1+H_{\rm BH}/2}$ \\
  $\beta^i_{\rm BH}$&$2\alpha_{\rm BH}^2 H_{\rm BH} l^i$&0 \\
  $\psi_{\rm BH}$&1&$ 1+H_{\rm BH}/2$ \\
  $\tilde{\gamma}_{ij}$&$\eta_{ij}+2H_{\rm BH} l_i l_j$&
  $\eta_{ij}$ \\ \hline
\end{tabular}
\end{center}
\label{table:bh}
\end{table}

In Table \ref{table:bh} we list the background metric quantities for
both Kerr-Schild (KS) and isotropic (ISO) coordinates \cite{footnote}.
 From these metric quantities a number of background quantities --
for example $\tilde R^i_{j}$ and $K$ -- that enter equations
(\ref{eq:lognp}), (\ref{eq:logn}) and (\ref{eq:shift}) are derived.
For Kerr-Schild coordinates these quantities can be found in BSS;
for isotropic coordinates they are either zero or trivial, since the
background metric $\tilde \gamma_{ij}$ is flat and the shift vanishes.

To aid in the numerical solution we decompose the metric quantities
into contributions from the neutron star and the black hole as 
follows.
The total lapse function and conformal factor are decomposed into the
product of a neutron star part and a black hole part (not a sum as in
BSS), such that
\begin{eqnarray}
  \alpha &=& \alpha_{\rm NS} \alpha_{\rm BH}, \label{eq:dec_lapse} \\
  \psi &=& \psi_{\rm NS} \psi_{\rm BH}. \label{eq:dec_conf}
\end{eqnarray}
The neutron star part is calculated by solving Poisson-like equations
while the black hole part is given by the background metric, shown in
Table~\ref{table:bh}.
The product form is required in order to
decompose $\nu$ and $\sigma$, defined by
Eqs. (\ref{eq:nu}) and (\ref{eq:sigma}), as a sum of the neutron star and
black hole parts, such that
\begin{eqnarray}
  \nu &=& \nu_{\rm NS} + \nu_{\rm BH}, \label{eq:dec_nu} \\
  \sigma &=& \sigma_{\rm NS} + \sigma_{\rm BH}, \label{eq:dec_sigma}
\end{eqnarray}
where $\nu_{\rm NS} \equiv \ln \alpha_{\rm NS}$,
$\nu_{\rm BH} \equiv \ln \alpha_{\rm BH}$,
$\sigma_{\rm NS} \equiv \ln (\alpha_{\rm NS} \psi_{\rm NS}^2)$,
and $\sigma_{\rm BH} \equiv \ln (\alpha_{\rm BH} \psi_{\rm BH}^2)$.
The decomposition (\ref{eq:dec_lapse}) and (\ref{eq:dec_conf}) are
different from those employed by BSS, but formally equivalent.

For the shift vector, we decompose as
\begin{equation}
  \beta^i = \beta^i_{\rm NS} +\beta^i_{\rm BH} +\beta^i_{\rm rot},
  \label{eq:dec_shift}
\end{equation}
where $\beta^i_{\rm NS}$ and $\beta^i_{\rm BH}$ are the neutron star and
black hole contributions to 
the shift vectors seen by the inertial observer, and
$\beta^i_{\rm rot}$ is the rotating shift vector defined as
\begin{equation}
  \beta^i_{\rm rot} = \epsilon^{ijk} \Omega_j X_k = \Omega (-Y,~X,~0),
\end{equation}
where $\Omega_j$ is the orbital angular velocity vector.  We define
the $Z$-axis to be parallel to the rotation axis, so that
$\Omega^i=(0,0,\Omega)$ in Cartesian coordinates.  

\begin{figure}[ht]
\begin{center}
  \includegraphics[width=8cm]{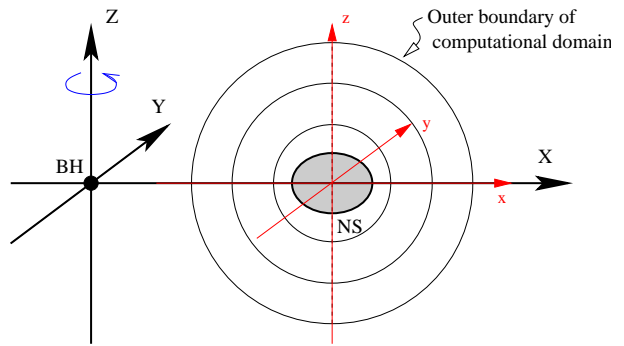}
\end{center}
\caption{Coordinate systems. The origin of the ``global'' coordinates
$(X,Y,Z)$ is located at the center of the black hole,
and that of the ``local'' coordinates $(x,y,z)$
at the point of maximum density within the neutron star.
Here, we assume an extreme mass ratio $M_{\rm BH} \gg M_{\rm NS}$,
so that the rotation axis of the orbit coincides with the center of the
black hole.
We define the $Z$-axis to point in the direction of the rotation axis.
The relation between the global and local coordinates is given by
$X=x+X_{\rm NS}$, $Y=y$, and $Z=z$,
where $X_{\rm NS}$ denotes the orbital separation between the black hole
and the neutron star.}
\label{fig:coord}
\end{figure}

As in BSS, we assume extreme mass ratios
$M_{\rm BH} \gg M_{\rm NS}$, which simplifies the problem in several ways.
In this limit, we may assume that the neutron star affects the space-time
only in a region around the neutron star itself, so that we can restrict
the computational domain to a neighborhood of the neutron star.
This means that we do not need excise a black hole singularity from the
numerical grid.
We may also assume that the rotation axis coincides with the center of
the black hole, which eliminates the need for an iteration to locate
the axis of rotation (see Fig. \ref{fig:coord}).

For the isotropic background, the solution is symmetric both across
the equatorial plane (i.e.~the $X\!\!-\!\!Y$ plane) and
the $X\!\!-\!\!Z$ plane.
As discussed in BSS and \cite{YoCBS04}, the presence of a
non-vanishing trace of the extrinsic curvature eliminates the symmetry
across the $X\!\!-\!\!Z$ plane for a Kerr-Schild background,
so that in the latter case we can assume a symmetry only across the
equatorial plane.

Evidently, the numerical implementation of the isotropic background is
much easier than that of the Kerr-Schild background.
However, isotropic coordinates have the disadvantage that the lapse
$\alpha_{\rm BH}$ vanishes on the black hole horizon, which would
cause problems in (\ref{eq:ext_curv_2}) when the black hole is
included in the computational domain.
The same property also causes problems in dynamical simulations,
when these quasiequilibrium models are used as initial data
(thus, in \cite{FaberBSTR05}, Poisson-like equations are solved for $\psi$
and $\alpha \psi$, as in BSS).
Kerr-Schild coordinates have the advantage that the coordinates smoothly
extend into the black hole interior, which eliminates both of these problems.

The above decompositions are then inserted into equations
(\ref{eq:lognp}), (\ref{eq:logn}) and (\ref{eq:shift}), which are then
solved for the neutron star contributions.
We list the resulting equations in Appendix \ref{app:eq_detail}.

\subsection{Hydrostatic equations}

For the cases considered here, the equations of relativistic
hydrodynamics reduce to the the integrated Euler equation as well as
the equation of continuity (which is satisfied identically for
corotating stars).
There are many sets of notation used to describe this
formulation \cite{BonaGM97,Asada97,Shiba98,Teuko98,footnote2},
but we follow that of \cite{GourGTMB01}, with one important
exception: we denote the adiabatic index of our polytropic EOS by
$\Gamma$ and the Lorentz factor of the matter by $\gamma$, whereas
their notation defines them the other way around.

\subsubsection{Integrated Euler equation}

The first integral of the Euler equation, common to both corotating
and irrotational configurations, can be written as
\begin{equation}
  h \alpha {\gamma \over \gamma_0} = {\rm constant}, \label{eq:i_euler}
\end{equation}
where $h=(\rho_0+\rho_i+P)/\rho_0$ is the fluid specific enthalpy.
Here $\gamma$ is the Lorentz factor between the fluid and the
co-orbiting observer and $\gamma_0$ that between the co-orbiting
observer and the inertial frame.
Here we define the "inertial" frame as the frame corresponding to
normal observers for whom the shift vector goes to zero at large
$r_{\rm BH}$. For the co-orbiting frame, normal observers satisfy
$\beta^i \rightarrow (\Omega \times r_{\rm BH})^i$ at large $r_{\rm BH}$.
A detailed derivation can be found, for example, in \cite{GourGTMB01}.
The Lorentz factors can be written as
\begin{eqnarray}
  \gamma &=& \gamma_{\rm n} \gamma_0 (1 -\gamma_{ij} U^i U_0^j), \\
  \gamma_0 &=& (1 -\gamma_{ij} U_0^i U_0^j)^{-1/2}, \label{eq:gamma0} \\
  \gamma_{\rm n} &=& (1 -\gamma_{ij} U^i U^j)^{-1/2},
\end{eqnarray}
where $U_0^i$ is the orbital 3-velocity with respect to the inertial
observer, given by $U_0^i=\beta^i/\alpha$, where
the shift vector $\beta^i$ is measured by a co-orbiting
observer.
The quantity $\gamma_{\rm n}$ denotes the Lorentz factor between the fluid
and the inertial observer, and can also be expressed as
$\gamma_{\rm n}=\alpha u^t$ where $u^t$ is the time component of
the fluid 4-velocity $u^{\mu}$.
The quantity $U^i$ is the fluid 3-velocity with
respect to the inertial observer (see Eq.~(27) of \cite{GourGTMB01}); 
for corotating binary systems, the fluid 3-velocity seen by the
co-orbiting observer vanishes, we obtain $U^i=U_0^i$.
For irrotational binary systems, $U^i$ can expressed in
terms of a velocity potential $\Psi$ as
\begin{equation}
  U^i ={1 \over \gamma_{\rm n} h} D^i \Psi,
\end{equation}
where $D_i$ is the covariant derivative with respect to $\gamma_{ij}$.
In this case, $\gamma_{\rm n}$ is written as
\begin{equation}
  \gamma_{\rm n} =\Bigl( 1 +{\gamma^{ij} D_i \Psi D_j \Psi \over h^2}
  \Bigr)^{1/2}.
\end{equation}

Taking the logarithm of Eq. (\ref{eq:i_euler}), we obtain the final form
of the integrated Euler equation,
\begin{equation}
  H_{\rm ent} +\nu -\ln \gamma_0 + \ln \gamma = {\rm constant},
  \label{eq:iEuler}
\end{equation}
where $H_{\rm ent}\equiv \ln h$.
For corotating binary systems we have
$\gamma=1$ and hence $\ln \gamma = 0$.

\subsubsection{Equation of continuity}

Having taken into account the helical symmetry,
we rewrite the equation of continuity
\begin{equation}
  {n \over h}  \nabla^{\mu} \nabla_{\mu} \Psi + (\nabla^{\mu} \Psi)
  \nabla_{\mu} \Bigl( {n \over h} \Bigr) = 0,
\end{equation}
where $n$ is the fluid baryon number density and $\nabla_{\mu}$ the
covariant derivative associated with $g_{\mu \nu}$, in the 3+1 form
\begin{eqnarray}
  &&n D^i D_i \Psi +(D^i n) (D_i \Psi) =h \gamma_{\rm n} U_0^i D_i n
  \nonumber \\
  &&+ n \Bigl[ (D^i \Psi) D_i \Bigl( \ln {h \over \alpha} \Bigr)
    +h U_0^i D_i \gamma_{\rm n} \Bigr] +n h K \gamma_{\rm n}.
  \label{eq:eoc_first}
\end{eqnarray}

Introducing an auxiliary quantity which represents
the logarithmic derivative of enthalpy with respect to
baryon number density,
\begin{equation}
  \zeta \equiv {d \ln H_{\rm ent} \over d \ln n},
\end{equation}
we can rewrite Eq. (\ref{eq:eoc_first}) as
\begin{eqnarray}
  &&\zeta H_{\rm ent} \tilde{D}^2 \Psi
  +\tilde{\gamma}^{ij} \tilde{D}_j H_{\rm ent} \tilde{D}_i \Psi
  =\psi^4 h \gamma_{\rm n} U_0^i \tilde{D}_i H_{\rm ent} \nonumber \\
  &&+\zeta H_{\rm ent} \Bigl[ \tilde{\gamma}^{ij} \tilde{D}_j \Psi
    \tilde{D}_i (H_{\rm ent} -\sigma) + \psi^4 h U_0^i \tilde{D}_i
    \gamma_{\rm n} \nonumber \\
  &&\hspace{40pt}+ \psi^4 h K \gamma_{\rm n} \Bigr]. \label{eq:eoc_t}
\end{eqnarray}
We will show the final form of equations for the cases of the Kerr-Schild
and isotropic coordinates in Appendices \ref{app:eoc_ks}
and \ref{app:eoc_cf}.

\subsection{Equation of state}

We adopt a polytropic EOS for the
neutron star of the form
\begin{equation}
  P =\kappa \rho_0^{\Gamma},
\end{equation}
where $\Gamma$ denotes the adiabatic index and $\kappa$ is a constant.
All results shown in this paper are for $\Gamma = 2$.
Since dimensions enter the problem only through $\kappa$, it is convenient
to rescale all dimensional quantities with respect to the length scale
\begin{equation}
  R_{\rm poly} \equiv \kappa^{1/2(\Gamma-1)}. \label{eq:poly_unit}
\end{equation}

For later comparison with the results of BSS,
we also introduce the dimensionless quantity
\begin{equation}
  q \equiv {P \over \rho_0},
\end{equation}
in terms of which we can express the baryon rest-mass density
and the pressure as
\begin{eqnarray}
  \rho_0 &=& R_{\rm poly}^{-2} q^{1/(\Gamma-1)}, \\
  P &=& R_{\rm poly}^{-2} q^{\Gamma/(\Gamma-1)}.
\end{eqnarray}

\subsection{Boundary condition} \label{sec:boundary}

For any Poisson-like elliptic equation of the form
\begin{equation}
  \underline{\Delta} \Phi = s,
\end{equation}
where $s$ is a source term with compact support, we can express the
exterior solution as
\begin{equation}
  \Phi (r, \theta, \varphi) =\sum_{l=0}^{\infty} \sum_{m=-l}^{l}
  s_{lm} Y_{lm} (\theta, \varphi) r^{-(l+1)}, \label{eq:poisson-solution}
\end{equation}
Here, $s_{lm}$ are the multipole moments, defined as
\begin{equation}
  s_{lm} \equiv \int_0^{R_{\rm B}} s r^l Y_{lm}^* (\theta, \varphi)
  d^3 x,
\end{equation}
and where $R_{\rm B}$ is the extent of the non-zero domain of
the source term.
The source terms of the Poisson-like equations (\ref{eq:logn_ks}) --
(\ref{eq:shift_ks}) or (\ref{eq:logn_cf}) -- (\ref{eq:shift_cf})
are not compact in the situation we consider;
however, since they fall off with a steep power-law dependence on the
radius, ignoring the contribution from outside the computational
domain introduces only minimal errors.
In our code we match the numerical solution to
(\ref{eq:poisson-solution}) at the outer boundaries, which we refer
to as a multipole boundary condition.

We truncate the expansion of the solution at a pre-determined value
$l=l_{\rm max}$, here setting $l_{\rm max}=4$ throughout.
We set $s_{lm}=0$ for all terms with $l>l_{\rm max}$.

Multipole boundary conditions, which allow us to achieve much greater
accuracy, are convenient in spectral applications,
but less so in finite difference implementations.
BSS, for example, used Robin boundary conditions which enforce that
fields fall of with a certain power of $1/r$.
In Sec. \ref{sec:type_bc}, we will compare
the results of these two boundary conditions (and we also refer to
\cite{FaberBSTR05} for a more detailed discussion of the boundary conditions).

\subsection{Determination of the orbital angular velocity}

To determine the orbital angular velocity, we require a force
balance along the $X$-axis at the center of the neutron star,
\begin{equation}
  {\partial H_{\rm ent} \over \partial X} \Bigl|_{(X_{\rm NS},0,0)} =0,
  \label{eq:fbalance}
\end{equation}
where $X_{\rm NS}$ is the $X$-coordinate of the center of the
neutron star relative to the black hole.
Equation (\ref{eq:fbalance}) means that the neutron star has its
maximum enthalpy (and thus density) at the position $(X_{\rm NS},0,0)$,
which corresponds to the origin of the local coordinate system $(x,y,z)$
in which we solve the equations
(see Fig.~\ref{fig:coord}).

Inserting Eq.~(\ref{eq:iEuler}) into the condition (\ref{eq:fbalance}),
we obtain
\begin{equation}
  {\partial \over \partial X} \ln \gamma_0 \Bigl|_{(X_{\rm NS},0,0)}
  = {\partial \over \partial X} (\nu +\ln \gamma) \Bigl|_{(X_{\rm NS},0,0)}.
   \label{eq:orbit}
\end{equation}
Equation (\ref{eq:orbit}) is solved algebraically for the orbital
angular velocity $\Omega$, which enters through the Lorentz factors
and the shift decomposition (\ref{eq:dec_shift})
on both sides of the equation.
We leave the dependence implicit on the right-hand side, but write out
the dependence explicitly on the left-hand side
(see Appendices \ref{app:ome_ks} and \ref{app:ome_cf}).

\subsection{Global Integrals}

It is reasonable to assume that during the binary inspiral both the
neutron star's rest-mass and the black hole's irreducible mass are
conserved.
The rest mass is defined as
\begin{eqnarray}
  M_{\rm NS} &=& \int \rho_0 u^t \sqrt{-g} d^3 x, \\
  &=& \int \rho_0 \alpha u^t \psi^6 \sqrt{\tilde{\gamma}} d^3 x,
\end{eqnarray}
where $g$ is the determinant of $g_{\mu \nu}$
and $\tilde{\gamma}$ that of $\tilde{\gamma}_{ij}$.
The determinant $\tilde{\gamma}$ takes the form
\begin{equation}
  \tilde{\gamma} = \left\{
\begin{array}{ll}
  1 +2H_{\rm BH} & ~~~{\rm for~the~K\!\!-\!\!S~background,} \\
  1 & ~~~{\rm for~the~flat~background.}
\end{array}
  \right.
\end{equation}
In polytropic units (see Eq. (\ref{eq:poly_unit})), we can normalize the
baryon rest-mass in dimensionless form
\begin{equation}
  \bar{M}_{\rm NS} \equiv {M_{\rm NS} \over R_{\rm poly}}.
\end{equation}

Assuming extreme mass ratios allows us to neglect tidal effects of the
neutron star on the black hole and to restrict the computational domain
to a neighborhood of the neutron star.
This means that we cannot evaluate the black hole's irreducible mass
$M_{\rm irr}$.
Instead, we keep the background mass $M_{\rm BH}$ constant in this paper.
The difference between $M_{\rm irr}$ and $M_{\rm BH}$ is of the order of
the binary's binding energy (see footnote [35] in BSS),
which is much smaller than either $M_{\rm irr}$ or $M_{\rm BH}$ in the
limit of extreme mass ratios.

Other global integrals that we might be interested in are the ADM mass
and the angular momentum, and especially their change along
equilibrium sequences as a function of the binary separation.
However, for the same reasons as explained in the previous paragraph,
we cannot capture contributions of the binding energy to these
quantities, meaning that we would find errors as large as the
quantities that we are interested in.
We therefore postpone evaluation of these integrals until we have relaxed
the assumption of extreme mass ratios.

\begin{figure}[ht]
\vspace{0.5cm}
\begin{center}
  \includegraphics[width=8cm]{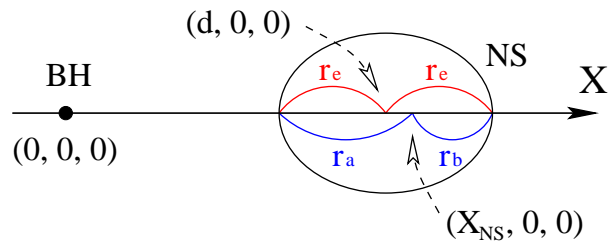}
\end{center}
\caption{Relation among the various neutron star ``radii'' $r_e$, $r_a$ and
$r_b$, as well as between the ``orbital separations'' $X_{\rm NS}$ and $d$.
$r_e$ is the half-diameter, which is defined as $r_e\equiv (r_a+r_b)/2$.}
\label{fig:radius}
\end{figure}

\subsection{Solution procedure}

We construct quasiequilibrium BHNS binaries in an iterative algorithm
as follows.
We start by preparing an initial guess, and then construct self-consistent
solutions to the equations listed in Appendices \ref{app:eq_ks} and
\ref{app:eq_cf} in an iteration that is similar to that of
\cite{GourGTMB01}.
Since there are some differences in various steps of the process, however,
it may be useful to describe the iteration in detail:
\begin{enumerate}
  \item Preparation of initial data for the main iteration

  \begin{enumerate}
    \item Construct a spherical star in equilibrium.

    \item Set the orbital separation between the center of the neutron star
      and the black hole to $X_{\rm NS}$.
      Here, the center of the neutron star is located at the maximum of
      the enthalpy (and thus the density).
      As we assume an extreme mass ratio, the absolute coordinate of the
      center of the black hole is $(0, 0, 0)$, that of the neutron star
      $(X_{\rm NS}, 0, 0)$; the rotation axis points in the $Z$-direction,
      and goes through the origin of the absolute coordinate
      (see Fig.~\ref{fig:coord}).
      At this stage, we set the black hole mass to a given constant
      $M_{\rm BH}$
      and never change it during the main iteration that follows.
  \end{enumerate}

  \item Main computation

  \begin{enumerate}
    \item Introduce the black hole gravitational field as a fixed
      background.

    \item Set the initial guess of the orbital angular velocity to
      the Keplerian frequency.

    \item Solve the set of equations \label{proc:solve}

    \begin{enumerate}
      \item Gravitational field equations: Eqs.~(\ref{eq:logn_ks}) --
	(\ref{eq:shift_ks}) for Kerr-Schild background;
	Eqs.~(\ref{eq:logn_cf}) -- (\ref{eq:shift_cf})
	for isotropic background
      \item Equation of continuity: Eq.~(\ref{eq:conti_ks}) for
	Kerr-Schild background;
	Eq.~(\ref{eq:conti_cf}) for isotropic background
    \end{enumerate}

    \item Determine the new enthalpy field from the integrated Euler
      equation, Eq.~(\ref{eq:iEuler}), and the new baryon rest-mass density
      from the neutron star EOS.

    \item Calculate the orbital angular velocity from Eq.~(\ref{eq:orbit})
      by solving it algebraically. \label{proc:orbit}

    \item Re-scale the radius of the neutron star and fit the outer
      boundary of the innermost domain to the surface of the neutron star.
      \label{proc:rescale}

    \item Change the orbital separation to have the same ratio
      of $X_{\rm NS}/r_e$ as the initially given value.
      Here $r_e$ is the half-diameter of the neutron star on the $X$-axis
      (see Fig.~\ref{fig:radius}). \label{proc:separation}

    \item Change the central enthalpy to fix the baryon rest-mass
      of the neutron star at a given value $M_{\rm NS}$.

    \item Compare the new enthalpy field with that of old one,
      and check whether the relative difference is smaller than the
      threshold or not. \label{proc:thres}

    \item If the condition (\ref{proc:thres}) is not satisfied,
      go back to (\ref{proc:solve}) and continue.
  \end{enumerate}
\end{enumerate}

The lack of a symmetry across the $X\!\!-\!\!Z$  plane for a
Kerr-Schild background introduces one more subtlety.
The procedure we introduced above determines all the eigenvalues
which appear in the present case, i.e., the orbital angular velocity,
the radius of the neutron star, and the integration constant of
the Euler equation.
However, the above procedure does not include a method to fix
the position of the neutron star in the local coordinate system $(x, y, z)$,
in which we solve the field equations.
Thanks to the $X\!\!-\!\!Y$ plane symmetry (equatorial symmetry),
the center of the neutron star is automatically located in the
$X\!\!-\!\!Y$ plane. Additionally, the procedure for determination
of the orbital angular velocity (\ref{proc:orbit}) fixes the local maximum
of the neutron star rest-mass density along the $X$-axis at $X=X_{\rm NS}$
($x=0$).
Lacking an additional constraint, the neutron star center can fall at
any position in the $Y$-direction, so long as the $X$-coordinate takes
the proper value.
To define our configurations unambiguously, we must require that the
position we impose in the procedure (\ref{proc:orbit}) is the {\it global}
maximum of the neutron star rest-mass density.
This requires the $Y$-derivative of the enthalpy to be zero as well as
the $X$-derivative at the point $(X_{\rm NS}, 0, 0)$.
To do so, we first introduce a function $f(y) = 1 - A y$,
where $A$ is a constant
defined by
\begin{equation}
  A \equiv \Bigl( {1 \over H_{\rm ent}}
  {\partial H_{\rm ent} \over \partial y} \Bigr)
  \Bigl|_{(X_{\rm NS}, 0, 0)}.
\end{equation}
When we multiply this function by the enthalpy and define a modified
enthalpy $H_{\rm mod} \equiv fH_{\rm ent}$, the modified enthalpy has
its global maximum at the position $(X_{\rm NS}, 0, 0)$.
During the iteration, this modified enthalpy term drags the neutron star
to the proper position in the $Y$-direction.
When the enthalpy maximum is properly located on the $X$-axis,
we recover $A=0$ and hence $f=1$.
We insert this procedure between (\ref{proc:rescale}) and
(\ref{proc:separation}) for the case of the Kerr-Schild background metric.

On the other hand, we have a $X\!\!-\!\!Z$ symmetry plane when using
the isotropic background.
Thus, the $Y$-derivative of the enthalpy always becomes zero on the
$X$-axis, and the global maximum of the enthalpy must fall on the $X$-axis.

\section{Code tests}

Our numerical code is based on the spectral methods libraries
developed by the Meudon relativity group \cite{Lorene}, and have been
tested, used, and found to be highly accurate in several previous
applications involving binary neutron stars
\cite{GourGTMB01,TanigGB01,TanigG02a,TanigG02b,TanigG03,BejgGGHTZ05}.
In the following we present some tests that verify our code for BHNS
binaries.
We present self-consistency tests, in which we explore the
sensitivity of the results to changes in the position of the outer
boundary, the type of boundary condition, and the number of
collocation points used in the spectral method, and compare
with analytical and previous numerical results.

\subsection{Self-consistency tests}

We present here the results of the self-consistency checks
for a neutron star mass $\bar{M}_{\rm NS}=0.05$.
Our convergence criteria are chosen so that
we iterate our solver until the relative difference in mass
between the given value and our numerical result is
less than one part in $10^{-7}$.
In the following tests, we will present values of the results
with five digits, although this overstates their true accuracy, in
order to show the magnitude of the changes in these quantities.

\subsubsection{Position of the outer boundary} \label{sec:position}

\begin{table*}[ht]
\caption{Convergence test for the position of the outer boundary $R_B$.}
\begin{center}
\begin{tabular}{c|ccc|ccc} \hline\hline
  \multicolumn{7}{c}
	      {Kerr-Schild: $\bar{M}_{\rm NS}=0.05$,
		$X_{\rm NS}/r_e=10.0$, $\bar{r}_0=1.1289$} \\ \hline\hline
  &\multicolumn{3}{c|}{Irrotation}&\multicolumn{3}{c}{Corotation} \\
  $R_{\rm B}/r_0$&$\Omega M_{\rm BH}$&$\bar{r}_e$&$q_{\rm max}$&
  $\Omega M_{\rm BH}$&$\bar{r}_e$&$q_{\rm max}$ \\ \hline
  4&9.7328(-3) &1.0964 &0.023493 &9.6399(-3) &1.1030 &0.023339 \\
  5&9.7432(-3) &1.0962 &0.023493 &9.6500(-3) &1.1029 &0.023338 \\
  6&9.7515(-3) &1.0962 &0.023493 &9.6580(-3) &1.1029 &0.023338 \\
  7&9.7603(-3) &1.0962 &0.023493 &9.6664(-3) &1.1029 &0.023338 \\
  8&9.7760(-3) &1.0962 &0.023493 &9.6807(-3) &1.1029 &0.023337 \\
  9&9.9358(-3) &1.0964 &0.023488 &9.7890(-3) &1.1032 &0.023330 \\ \hline\hline
  \multicolumn{7}{c}
	      {Isotropic: $\bar{M}_{\rm NS}=0.05$,
		$X_{\rm NS}/r_e=10.0$, $\bar{r}_0=1.1289$} \\ \hline\hline
  &\multicolumn{3}{c|}{Irrotation}&\multicolumn{3}{c}{Corotation} \\
  $R_{\rm B}/r_0$&$\Omega M_{\rm BH}$&$\bar{r}_e$&$q_{\rm max}$&
  $\Omega M_{\rm BH}$&$\bar{r}_e$&$q_{\rm max}$ \\ \hline
  4&9.1059(-3) &1.0960 &0.023507 &9.0364(-3) &1.1017 &0.023376 \\
  5&9.1093(-3) &1.0960 &0.023507 &9.0397(-3) &1.1017 &0.023376 \\
  6&9.1114(-3) &1.0960 &0.023507 &9.0418(-3) &1.1017 &0.023376 \\
  7&9.1129(-3) &1.0960 &0.023507 &9.0432(-3) &1.1017 &0.023376 \\
  8&9.1139(-3) &1.0960 &0.023507 &9.0443(-3) &1.1017 &0.023376 \\
  9&9.1149(-3) &1.0960 &0.023507 &9.0452(-3) &1.1017 &0.023376 \\ \hline
\end{tabular}
\end{center}
\label{table:position_ob}
\end{table*}

As long as the outer boundary of our computational domain is
sufficiently far away from both the neutron star and the black hole,
any physical quantities should be unaffected by the exact location of
the outer boundary.
Note that the outer boundary is located between the neutron star
and the black hole in the present work (see Fig.~\ref{fig:coord}).
To test this, we set the orbital separation to $X_{\rm NS}/r_e=10$,
and change the position of the outer boundary from
$R_{\rm B}/r_0=4$ to 9, where $r_0$ denotes the radius of a spherical
neutron star with the same baryon rest-mass.
In Table \ref{table:position_ob}, we show the
orbital angular velocity $\Omega M_{\rm BH}$, the half-diameter
of the neutron star $\bar{r}_e \equiv r_e/R_{\rm poly}$, and the maximum
of the density quantity $q_{\rm max}$.
One can see from Table \ref{table:position_ob} that the radius of the
neutron star and the maximum of the density quantity
converge to better than order $10^{-5}$ for $5 <R_{\rm B}/r_0 \le 8$,
while the convergence of the orbital angular velocity is only of order
$10^{-3}$ for a Kerr-Schild background and $10^{-4}$ for an isotropic
background.

For a separation $X_{\rm NS}/r_e=10$ and $\bar{M}_{\rm NS}=0.05$,
no position of the outer boundary larger than $R_{\rm B}/r_0 >9.7$ is
permitted because the boundary will overlap the black hole singularity.
Even for a smaller value, the source terms of
Eqs. (\ref{eq:logn_ks}) -- (\ref{eq:shift_ks})
or (\ref{eq:logn_cf}) -- (\ref{eq:shift_cf})
do not decrease sufficiently quickly with radius because of the contributions
from the black hole background metric.
Thus, we have to choose a position of the outer boundary
for each orbital separation whose position is neither close to the neutron
star nor to the black hole.

The relative error of the orbital angular velocity listed in
Table \ref{table:position_ob} is larger for the Kerr-Schild background than
for the isotropic background.
This tendency results in part from the absence of the $X\!\!-\!\!Z$ plane
symmetry in the Kerr-Schild background.
We also see that the orbital angular velocity determined by using a
larger outer boundary
radius is slightly larger than that found using a smaller one.
Because of this ambiguity, we will show only three significant digits
in the final results.

\subsubsection{Type of boundary condition} \label{sec:type_bc}

\begin{table*}[ht]
\caption{Convergence test for the type of boundary condition.}
\begin{center}
\begin{tabular}{ccc|ccc|ccc} \hline\hline
  \multicolumn{9}{c}
	      {Kerr-Schild: $\bar{M}_{\rm NS}=0.05$} \\ \hline\hline
  &&&\multicolumn{3}{c|}{Irrotation}&\multicolumn{3}{c}{Corotation} \\
  Boundary condition&$X_{\rm NS}/r_e$&$R_{\rm B}/r_0$&
  $\Omega M_{\rm BH}$&$\bar{r}_e$&$q_{\rm max}$&
  $\Omega M_{\rm BH}$&$\bar{r}_e$&$q_{\rm max}$ \\ \hline
  multipole &10.0&8&9.7760(-3) &1.0962 &0.023493 &
                           9.6807(-3) &1.1029 &0.023337 \\
  leading order    &10.0&8&9.7799(-3) &1.0962 &0.023493 &
			   9.6843(-3) &1.1029 &0.023337 \\
  multipole &8.0 &7&1.3588(-2) &1.1062 &0.023429 &
                           1.3325(-2) &1.1186 &0.023126 \\
  leading order    &8.0 &7&1.3627(-2) &1.1064 &0.023427 &
			   1.3350(-2) &1.1188 &0.023125 \\
  multipole &6.0 &5&1.9457(-2) &1.1583 &0.023193 &
                           1.8837(-2) &1.1818 &0.022558 \\
  leading order    &6.0 &5&1.9478(-2) &1.1584 &0.023192 &
			   1.8855(-2) &1.1817 &0.022558 \\ \hline\hline
  \multicolumn{9}{c}
	      {Isotropic: $\bar{M}_{\rm NS}=0.05$} \\ \hline\hline
  &&&\multicolumn{3}{c|}{Irrotation}&\multicolumn{3}{c}{Corotation} \\
  Boundary condition&$X_{\rm NS}/r_e$&$R_{\rm B}/r_0$&
  $\Omega M_{\rm BH}$&$\bar{r}_e$&$q_{\rm max}$&
  $\Omega M_{\rm BH}$&$\bar{r}_e$&$q_{\rm max}$ \\ \hline
  multipole &10.0&8&9.1139(-3) &1.0960 &0.023507 &
                           9.0443(-3) &1.1017 &0.023376 \\
  leading order    &10.0&8&9.1152(-3) &1.0960 &0.023507 &
			   9.0455(-3) &1.1017 &0.023376 \\
  multipole &8.0 &7&1.2430(-2) &1.1021 &0.023467 &
                           1.2261(-2) &1.1126 &0.023221 \\
  leading order    &8.0 &7&1.2432(-2) &1.1022 &0.023467 &
			   1.2263(-2) &1.1126 &0.023221 \\
  multipole &6.0 &5&1.7811(-2) &1.1381 &0.023315 &
                           1.7372(-2) &1.1580 &0.022800 \\
  leading order    &6.0 &5&1.7816(-2) &1.1382 &0.023315 &
			   1.7377(-2) &1.1581 &0.022800 \\ \hline
\end{tabular}
\end{center}
\label{table:type_bc}
\end{table*}

Next, we show in Table \ref{table:type_bc} the results of a comparison
between our two different boundary conditions.
The first is the multipole condition described in Sec.~\ref{sec:boundary},
and second is the condition in which only the leading-order, fall-off term
is taken for each metric component.
It is found that the relative difference in the orbital angular velocity
is smaller than  order $10^{-3}$ for the Kerr-Schild background and
$10^{-4}$ for the isotropic background.
The leading-order fall-off boundary condition seems to overestimate the
angular velocity relative to the multipole condition.
However, the relative error introduced by using different boundary conditions
is an order of magnitude
less than that introduced by changing the position of the outer boundary.

\subsubsection{Number of collocation points}

In the last self-consistency check we verify that physical quantities
are largely independent of the number of collocation points $N_r
\times N_{\theta} \times N_{\varphi}$ used in spectral field solver.
Here, $N_r$, $N_{\theta}$, and $N_{\varphi}$ denote the number of collocation
points in the radial, polar, and azimuthal directions, respectively.
The orbital angular velocity seems to converge at a level of $10^{-4}$
in the Kerr-Schild background and $< 10^{-5}$ in the isotropic background,
for any number of collocation points larger than $25 \times 17 \times 16$.
Since the error in the orbital angular velocity is smaller than that
induced by changing the position of the outer boundary for this number of
points, we use a fixed number of collocation points
$N_r \times N_{\theta} \times N_{\varphi} =25 \times 17 \times 16$
to save computational time.

\begin{table*}[ht]
\caption{Convergence test for the number of collocation points
used in the spectral method field solver,
$N_r \times N_{\theta} \times N_{\varphi}$.}
\begin{center}
\begin{tabular}{c|ccc|ccc} \hline\hline
  \multicolumn{7}{c}
	      {Kerr-Schild: $\bar{M}_{\rm NS}=0.05$,
		$X_{\rm NS}/r_e=10.0$, $R_{\rm B}/r_0=8$} \\ \hline\hline
  &\multicolumn{3}{c|}{Irrotation}&\multicolumn{3}{c}{Corotation} \\
  $N_r \times N_{\theta} \times N_{\varphi}$&
  $\Omega M_{\rm BH}$&$\bar{r}_e$&$q_{\rm max}$&
  $\Omega M_{\rm BH}$&$\bar{r}_e$&$q_{\rm max}$ \\ \hline
  $13 \times 9 \times 8$  &9.8004(-3) &1.0961 &0.023495 &
	                   9.7019(-3) &1.1030 &0.023338 \\
  $17 \times 13 \times 12$&9.7807(-3) &1.0962 &0.023493 &
			   9.6849(-3) &1.1029 &0.023337 \\
  $25 \times 17 \times 16$&9.7760(-3) &1.0962 &0.023493 &
			   9.6807(-3) &1.1029 &0.023337 \\
  $33 \times 21 \times 20$&9.7742(-3) &1.0962 &0.023493 &
			   9.6792(-3) &1.1029 &0.023337 \\
  $37 \times 25 \times 24$&9.7735(-3) &1.0962 &0.023493 &
			   9.6785(-3) &1.1029 &0.023337 \\ \hline\hline
  \multicolumn{7}{c}
	      {Isotropic: $\bar{M}_{\rm NS}=0.05$,
		$X_{\rm NS}/r_e=10.0$, $R_{\rm B}/r_0=8$} \\ \hline\hline
  &\multicolumn{3}{c|}{Irrotation}&\multicolumn{3}{c}{Corotation} \\
  $N_r \times N_{\theta} \times N_{\varphi}$&
  $\Omega M_{\rm BH}$&$\bar{r}_e$&$q_{\rm max}$&
  $\Omega M_{\rm BH}$&$\bar{r}_e$&$q_{\rm max}$ \\ \hline
  $13 \times 9 \times 8$  &9.1144(-3) &1.0959 &0.023511 &
	                   9.0468(-3) &1.1016 &0.023379 \\
  $17 \times 13 \times 12$&9.1140(-3) &1.0960 &0.023507 &
			   9.0443(-3) &1.1017 &0.023376 \\
  $25 \times 17 \times 16$&9.1139(-3) &1.0960 &0.023507 &
			   9.0443(-3) &1.1017 &0.023376 \\
  $33 \times 21 \times 20$&9.1139(-3) &1.0960 &0.023507 &
			   9.0443(-3) &1.1017 &0.023376 \\
  $37 \times 25 \times 24$&9.1139(-3) &1.0960 &0.023507 &
			   9.0443(-3) &1.1017 &0.023376 \\ \hline
\end{tabular}
\end{center}
\label{table:number_cp}
\end{table*}

\subsection{Comparison with previous results}

To date, previous attempts to model equilibrium BHNS binary sequences have
included the construction of
a corotating relativistic star in a black hole background
metric \cite{BaumSS04},
a corotating Newtonian star in a black hole background
metric computed in the tidal approximation \cite{IshiiSM05}, and
a corotating or irrotational BHNS binary system in Newtonian
gravity \cite{UryuE99}.
We now compare our relativistic results with these calculations.

\subsubsection{Comparison with BSS} \label{sec:comp_BSS04}

\begin{table}[ht]
\caption{Comparison with the results of BSS, which are calculated
in the Kerr-Schild background metric
for corotating BHNS binaries.
The neutron star has a baryon rest-mass $\bar{M}_{\rm NS}=0.05$,
and the mass ratio is $M_{\rm BH}/M_{\rm NS}=10$.}
\begin{center}
\begin{tabular}{lccccc} \hline\hline
  &$d/r_e$&$d/M_{\rm BH}$&$\Omega M_{\rm BH}$&$~~~\bar{r}_e~~~$&
  $~~q_{\rm max}~~$ \\ \hline
  BSS         &8.0 &17.8 &0.0133 &1.11 &0.0235 \\
  Our results &8.0 &17.9 &0.0133 &1.12 &0.0231 \\ \hline
  BSS         &5.0 &12.6 &0.0223 &1.26 &0.0223 \\
  Our results &5.0 &12.7 &0.0222 &1.28 &0.0219 \\ \hline
\end{tabular}
\end{center}
\label{table:comp_BSS04}
\end{table}

BSS provide results for a constant rest-mass sequence of corotating
BHNS binaries in a Kerr-Schild background, with $\bar{M}_{\rm NS}=0.05$
and $M_{\rm BH}/M_{\rm NS}=10$.
We compare our results at two of the orbital separations listed in their
Table I, $d/r_e=8$ and 5.
Here $d$ is the coordinate separation from the center of the black hole
to the half-diameter of the neutron star on the $X$-axis,
a quantity similar to their ``$-x_{\rm BH}$''.
We define 
\begin{equation}
  d \equiv X_{\rm NS} + {r_b - r_a \over 2},
\end{equation}
where $r_a$ is the radius of the neutron star in the direction toward
the black hole and $r_b$ that to the opposite side
(see Fig.~\ref{fig:radius}).
Since our definition of the orbital separation $X_{\rm NS}$
is different from that found in BSS, we have to transform
our separation $X_{\rm NS}$ to $d$.
However, since the relative difference between these two
separations is less than 1\% even for $d/r_e=5$,
our results are insensitive to the change up to three significant digits.

In Table \ref{table:comp_BSS04}, we present the results of the comparison.
Note that we show the results at $d/r_e=5$ in the table here, but
our results at this separation do not converge as well as our other results,
only to the level of several parts in $10^{-5}$ in the relative error of the
enthalpy between two successive iteration steps.
Thus we do not include those results
in Sec. \ref{sec:results}, but show them here as a code test only.

Our results agree with those of BSS within 2\%,
even though we calculate them using a completely different numerical method.

\subsubsection{Comparison with the results by Ishii {\it et al.}
\protect\cite{IshiiSM05}}

In \cite{IshiiSM05}, the critical value of a quantity
\begin{equation}
  \mu \equiv \Bigl( {M_{\rm NS} \over M_{\rm BH}} \Bigr)
  \Bigl( {X_{\rm NS} \over r_0} \Bigr)^3, \label{eq:mu_def}
\end{equation}
was calculated for corotating BHNS binaries in Fermi normal coordinates.
Here $r_0$ is the radius that a neutron star with the same mass would have
in isolation.
In these calculations, the neutron star is treated as a Newtonian star in
the black hole background which is expanded up to fourth order 
in the parameter $r_0/X_{\rm NS}$.
The quantity above is defined such that if $\mu < \mu_{\rm crit}$,
the neutron star will be tidally disrupted by the black hole.
By fitting their numerical results,
they give an approximate formula for $\mu_{\rm crit}$ which
holds for separations $X_{\rm NS} \ge 6 M_{\rm BH}$ with good accuracy.
The formula, Eq. (186) of \cite{IshiiSM05}, is given by
\begin{equation}
  \mu_{\rm crit} \simeq 14.9 \Bigl( 1 +0.80 {r_0 \over X_{\rm NS}} \Bigr)
  \label{eq:mu_crit}
\end{equation}
for polytropic neutron star models with an adiabatic index $\Gamma=2$.

\begin{figure}[ht]
\vspace{0.5cm}
\begin{center}
  \includegraphics[width=8cm]{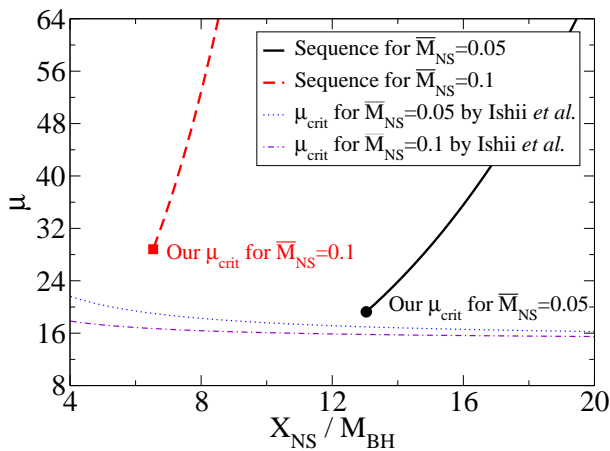}
 \end{center}
\caption{Critical value of the tidal disruption parameter $\mu$ as
a function of the orbital separation.
The thick solid and dashed lines represent sequences of corotating BHNS
binaries with neutron star masses $\bar{M}_{\rm NS}=0.05$ and 0.1,
respectively; the filled circle and square show the critical values for both.
The thin dotted and dot-dashed lines denote the critical values given by Eq.
(\ref{eq:mu_crit}) for neutron star masses $\bar{M}_{\rm NS}=0.05$ and
0.1.}
\label{fig:test_comp_ISM}
\end{figure}

We show the results of the comparison in Fig. \ref{fig:test_comp_ISM}.
The thick solid and dashed lines correspond to sequences of corotating
BHNS binaries with neutron star masses $\bar{M}_{\rm NS}=0.05$ and 0.1,
respectively, while the filled circle and square show the critical values
for these sequences.
These points are calculated by inserting the closest separation we obtained
for the corotating state in the Kerr-Schild background case
into Eq. (\ref{eq:mu_def}).
The thin dotted and dot-dashed lines denote the critical value given by
Eq. (\ref{eq:mu_crit}) for neutron star masses $\bar{M}_{\rm NS}=0.05$
and 0.1, where we use the radius of a spherical neutron star with
the same baryon rest-mass; $\bar{r}_0=1.1289$ for $\bar{M}_{\rm NS}=0.05$
and $\bar{r}_0=0.98972$ for $\bar{M}_{\rm NS}=0.1$.
for drawing all critical points.

Evidently, our sequences do not quite reach the limits found in
\cite{IshiiSM05}.
This may be because our spectral method is unable to treat the cusp-like
features that appear at the inner Lagrange point
when the neutron star fills out its Roche lobe.
We can therefore approach the tidal break-up separation only up to
a few percent, which may explain why our results for $\mu_{\rm crit}$
are somewhat larger than those found in \cite{IshiiSM05}.

\subsubsection{Comparison with the results
by Uryu \& Eriguchi \protect\cite{UryuE99}}

\begin{table}[ht]
\caption{Comparison with the results of Uryu and Eriguchi \cite{UryuE99}.
The results are compared for a neutron star mass
$\bar{M}_{\rm NS}=0.05$ with mass ratio $M_{\rm BH}/M_{\rm NS}=10$.
Here, we use the radius of a spherical star,
$\bar{r}_0=1.1289$, to convert their results into our units.
K-S and ISO denote the respective backgrounds.}
\begin{center}
\begin{tabular}{lc|c} \hline \hline
  &\multicolumn{1}{c|}{Irrotation}&\multicolumn{1}{c}{Corotation} \\
  &$\Omega M_{\rm BH}$&$\Omega M_{\rm BH}$ \\ \hline
  Uryu \& Eriguchi&0.0262 &0.0264 \\
  Our results (K-S)&0.0224 &0.0216 \\
  Our results (ISO)&0.0224 &0.0217 \\ \hline
\end{tabular}
\end{center}
\label{table:comp_UE99}
\end{table}

In this section, we compare our results with those of \cite{UryuE99},
as was done by BSS with good agreement.
In \cite{UryuE99} sequences of BHNS binaries in Newtonian gravity
were computed for both irrotational flow and corotation.
For comparison, we select data with $N=1.0$ and $M_{\rm S}/M_{\rm BH}=0.1$
in their Table 2 for
the corotating case, and in Table 4 for the irrotational case.
Here, $N$ denotes the polytropic index ($N\equiv 1/(\Gamma -1)$)
and $M_{\rm S}$ is
the mass of the star (corresponding to $M_{\rm NS}$ here).
We present the results of a comparison of the orbital angular velocity
in Table \ref{table:comp_UE99}.
There, we compare our data for a neutron star mass
$\bar{M}_{\rm NS}=0.05$ and mass ratio $M_{\rm BH}/M_{\rm NS}=10$
in the Kerr-Schild and isotropic backgrounds.
Since the sequences in \cite{UryuE99} were computed using Newtonian
gravity while we use full general relativity,
we can compare only an invariant quantity, e.g., the orbital angular velocity.
In order to convert their Newtonian results to values comparable
to our relativistic ones, we scale our results by the radius of a
relativistic spherical neutron star with mass $\bar{M}_{\rm NS}=0.05$,
$\bar{r}_0=1.1289$.
Our results are shown at the closest separation we calculate,
which is several percent further away in distance from
the real tidal disruption point,
because the spectral method is unable to treat cusp-like figures.
Thus we find our calculated values are about 15 \% smaller
than those found in \cite{UryuE99}, but given that our configurations
have separations several percent away from the real tidal disruption point,
we can say that the agreement with the angular velocities
in Table \ref{table:comp_UE99} is reasonable.

\subsection{Comparison with the relativistic Keplerian velocity in the
  limit of extreme mass ratio}

\begin{figure*}[ht]
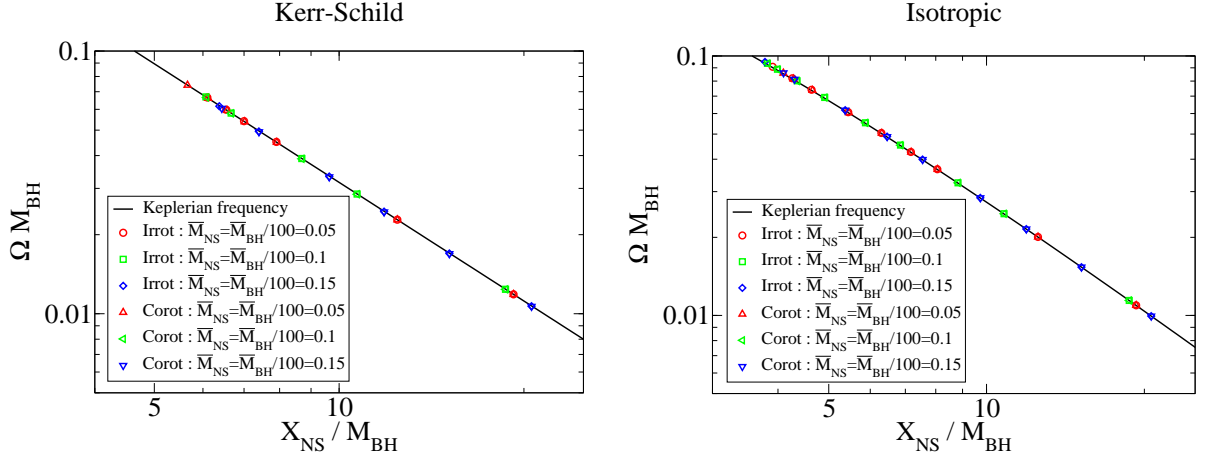

\vspace{0.5cm}
\begin{center}
  \includegraphics[height=6cm]{fig4a_kep_ks.eps}
  \hspace{10pt}
  \includegraphics[height=6cm]{fig4b_kep_cf.eps}
\end{center}
\caption{The orbital angular velocity as a function of the orbital separation
for a mass ratio $M_{\rm BH}/M_{\rm NS}=100$.
The left and right panels show Kerr-Schild and isotropic backgrounds,
respectively.
In each panel, we present six different sequences:
two choices for the rotational state (irrotation and corotation) for
each of three different neutron star masses
($\bar{M}_{\rm NS}=0.05$, 0.1, and 0.15).
The solid line in each panel denotes the frequency of a Keplerian orbit. 
Note that the horizontal axis shows the 
$X$ coordinate in Kerr-Schild coordinates in the left panel, but isotropic
coordinates on the right.}
\label{fig:test_kep}
\end{figure*}

In the limit of an extreme mass ratio ($M_{\rm BH}/M_{\rm NS} \gg 1$),
the neutron star orbits on a circular, test-particle geodesic.
In this limit, we should recover Kepler's law,
$\Omega=\sqrt{M_{\rm BH}/r^3}$,
where $r$ is the areal radius.
In Kerr-Schild coordinates, the radial coordinate is areal,
but in isotropic coordinates we have to transform according to
\begin{equation}
  r = (\psi_{\rm BH}^{\rm ISO})^2 r_{\rm ISO}
  = \Bigl( 1 +{M_{\rm BH} \over 2 r_{\rm ISO}} \Bigr)^2 r_{\rm ISO},
\end{equation}
where $r_{\rm ISO}$ denotes the isotropic radius.
In Fig.~\ref{fig:test_kep}, we present the results for
a mass ratio $M_{\rm BH}/M_{\rm NS}=100$
with neutron star masses of $\bar{M}_{\rm NS}=0.05$, 0.1, and 0.15.
It is obvious that our results are in good agreement
with the Keplerian frequency for both irrotational and corotating flows.

\section{Numerical results}

All results presented in this Section are for a polytropic index
$\Gamma = 2$.
The computational grid is divided into four domains
(see Fig. \ref{fig:coord}), each one of which is covered by 
$N_r \times N_{\theta} \times N_{\varphi}=25 \times 17 \times 16$
collocation points.
We focus on a mass ratio of $M_{\rm BH}/M_{\rm NS}=10$
and present 12 different sequences including
\begin{itemize}
  \item Kerr-Schild backgrounds and isotropic backgrounds

  \item irrotational and corotational fluid flow

  \item neutron star mass of $\bar{M}_{\rm NS}=0.05$, 0.1, and 0.15.
\end{itemize}
The results are summarized in Tables \ref{table:seq_K005} --
\ref{table:seq_K015} for Kerr-Schild backgrounds and
\ref{table:seq_C005} -- \ref{table:seq_C015} for isotropic
backgrounds.
Note that an isolated neutron star with the adopted $\Gamma=2$
polytropic EOS has a maximum mass $\bar{M}_{\rm NS,max}=0.18$.

\subsection{The Kerr-Schild background} \label{sec:results}

\begin{table*}[ht]
\caption{Physical parameters along a sequence with
$M_{\rm BH}/M_{\rm NS}=10$ in Kerr-Schild backgrounds.
The orbital separation $X_{\rm NS}$, orbital angular velocity $\Omega$,
half-diameter along the $X$-axis $\bar{r}_e$, maximum of
the density parameter $q_{\rm max}$, relative change in the central
energy density $\delta e$, mass-shedding indicator $\chi$, and radius of
the outer boundary $R_B$ are shown.
The baryon rest-mass of the neutron star is $\bar{M}_{\rm NS}=0.05$.
$M_{\rm NS}^G$ denotes the gravitational mass of a spherical neutron star
with the same baryon rest-mass.
$M_{\rm NS}^G/r_0$ is the compaction parameter of an isolated neutron star.}
\begin{center}
\begin{tabular}{rccccccc} \hline\hline
  \multicolumn{8}{c}{Kerr-Schild: $\bar{M}_{\rm NS}=0.05$,
    $M_{\rm NS}^G/r_0=0.0415$, $\bar{r}_0=1.1289$} \\ \hline\hline
  \multicolumn{8}{c}{Irrotation} \\
  $X_{\rm NS}/r_e$&$X_{\rm NS}/M_{\rm BH}$&$\Omega M_{\rm BH}$&
  $~~~\bar{r}_e~~~$&$~~q_{\rm max}~~$&$\delta e$&$~~~~\chi~~~~$&
  $R_{\rm B}/r_0$ \\ \hline
  10.0&21.9 &9.78(-3) &1.10 &0.0235 &-2.41(-3) &0.977 &8 \\
  8.0 &17.7 &1.36(-2) &1.11 &0.0234 &-5.16(-3) &0.908 &7 \\
  7.0 &15.7 &1.62(-2) &1.12 &0.0234 &-8.39(-3) &0.834 &6 \\
  6.0 &13.9 &1.95(-2) &1.16 &0.0232 &-1.54(-2) &0.699 &5 \\
  5.2 &12.8 &2.24(-2) &1.23 &0.0229 &-2.89(-2) &0.505 &5 \\ \hline\hline
  \multicolumn{8}{c}{Corotation} \\
  $X_{\rm NS}/r_e$&$X_{\rm NS}/M_{\rm BH}$&$\Omega M_{\rm BH}$&$\bar{r}_e$&
  $q_{\rm max}$&$\delta e$&$\chi$&$R_{\rm B}/r_0$ \\ \hline
  10.0&22.1 &9.68(-3) &1.10 &0.0233 &-9.17(-3) &0.957 &8 \\
  8.0 &17.9 &1.33(-2) &1.12 &0.0231 &-1.83(-2) &0.875 &7 \\
  7.0 &15.9 &1.58(-2) &1.14 &0.0229 &-2.73(-2) &0.793 &6 \\
  6.0 &14.2 &1.88(-2) &1.18 &0.0226 &-4.30(-2) &0.654 &5 \\
  5.2 &13.0 &2.16(-2) &1.25 &0.0220 &-6.51(-2) &0.460 &5 \\ \hline
\end{tabular}
\end{center}
\label{table:seq_K005}
\end{table*}

\begin{table*}[ht]
\caption{Same as Table \ref{table:seq_K005}, but for $\bar{M}_{\rm NS}=0.1$.}
\begin{center}
\begin{tabular}{rcccccccc} \hline\hline
  \multicolumn{8}{c}{Kerr-Schild: $\bar{M}_{\rm NS}=0.1$,
    $M_{\rm NS}^G/r_0=0.0879$, $\bar{r}_0=0.98972$} \\ \hline\hline
  \multicolumn{8}{c}{Irrotation} \\
  $X_{\rm NS}/r_e$&$X_{\rm NS}/M_{\rm BH}$&$\Omega M_{\rm BH}$&
  $~~~\bar{r}_e~~~$&$~~q_{\rm max}~~$&$\delta e$&$~~~~\chi~~~~$&
  $R_{\rm B}/r_0$ \\ \hline
  14.0&12.8 &2.19(-2) &0.915 &0.0585 &-6.69(-3) &1.06 &8 \\
  12.0&10.9 &2.82(-2) &0.905 &0.0582 &-1.10(-2) &1.05 &8 \\
  10.0&8.96 &3.78(-2) &0.896 &0.0577 &-2.06(-2) &1.02 &7 \\
  8.0 &7.20 &5.29(-2) &0.901 &0.0564 &-4.31(-2) &0.912 &5.5 \\
  7.4 &6.78 &5.88(-2) &0.916 &0.0554 &-6.16(-2) &0.821 &5.5 \\ \hline\hline
  \multicolumn{8}{c}{Corotation} \\
  $X_{\rm NS}/r_e$&$X_{\rm NS}/M_{\rm BH}$&$\Omega M_{\rm BH}$&
  $~~~\bar{r}_e~~~$&$~~q_{\rm max}~~$&$\delta e$&$~~~~\chi~~~~$&
  $R_{\rm B}/r_0$ \\ \hline
  14.0&12.8 &2.18(-2) &0.918 &0.0582 &-1.07(-2) &1.05 &8 \\
  12.0&10.9 &2.79(-2) &0.910 &0.0578 &-1.76(-2) &1.03 &8 \\
  10.0&9.06 &3.71(-2) &0.906 &0.0570 &-3.24(-2) &0.990&7 \\
  8.0 &7.36 &5.15(-2) &0.920 &0.0550 &-6.87(-2) &0.860&6 \\
  6.8 &6.54 &6.13(-2) &0.961 &0.0527 &-1.09(-1) &0.684&5 \\ \hline
\end{tabular}
\end{center}
\label{table:seq_K01}
\end{table*}

\begin{table*}[ht]
\caption{Same as Table \ref{table:seq_K005}, but for $\bar{M}_{\rm NS}=0.15$.}
\begin{center}
\begin{tabular}{rcccccccc} \hline\hline
  \multicolumn{8}{c}{Kerr-Schild: $\bar{M}_{\rm NS}=0.15$,
    $M_{\rm NS}^G/r_0=0.145$, $\bar{r}_0=0.81526$} \\ \hline\hline
  \multicolumn{8}{c}{Irrotation} \\
  $X_{\rm NS}/r_e$&$X_{\rm NS}/M_{\rm BH}$&$\Omega M_{\rm BH}$&
  $~~~\bar{r}_e~~~$&$~~q_{\rm max}~~$&$\delta e$&$~~~~\chi~~~~$&
  $R_{\rm B}/r_0$ \\ \hline
  22.0 &10.7 &2.84(-2) &0.733 &0.126 &-5.76(-3) &1.10 &8 \\
  20.0 &9.65 &3.34(-2) &0.724 &0.126 &-8.32(-3) &1.11 &8 \\
  18.0 &8.55 &4.02(-2) &0.712 &0.125 &-1.26(-2) &1.13 &8 \\
  16.0 &7.45 &4.97(-2) &0.698 &0.124 &-2.09(-2) &1.15 &8 \\
  14.4 &6.57 &6.05(-2) &0.683 &0.123 &-3.04(-2) &1.16 &8 \\ \hline\hline
  \multicolumn{8}{c}{Corotation} \\
  $X_{\rm NS}/r_e$&$X_{\rm NS}/M_{\rm BH}$&$\Omega M_{\rm BH}$&
  $~~~\bar{r}_e~~~$&$~~q_{\rm max}~~$&$\delta e$&$~~~~\chi~~~~$&
  $R_{\rm B}/r_0$ \\ \hline
  22.0 &10.8 &2.83(-2) &0.734 &0.126 &-7.55(-3) &1.10 &8 \\
  20.0 &9.67 &3.32(-2) &0.725 &0.125 &-1.09(-2) &1.11 &8 \\
  18.0 &8.58 &3.99(-2) &0.715 &0.125 &-1.64(-2) &1.12 &8 \\
  16.0 &7.49 &4.91(-2) &0.702 &0.124 &-2.68(-2) &1.13 &8 \\
  14.0 &6.41 &6.27(-2) &0.686 &0.121 &-4.86(-2) &1.14 &8 \\ \hline
\end{tabular}
\end{center}
\label{table:seq_K015}
\end{table*}

\begin{figure*}[ht]
\begin{center}
  \includegraphics[height=7cm]{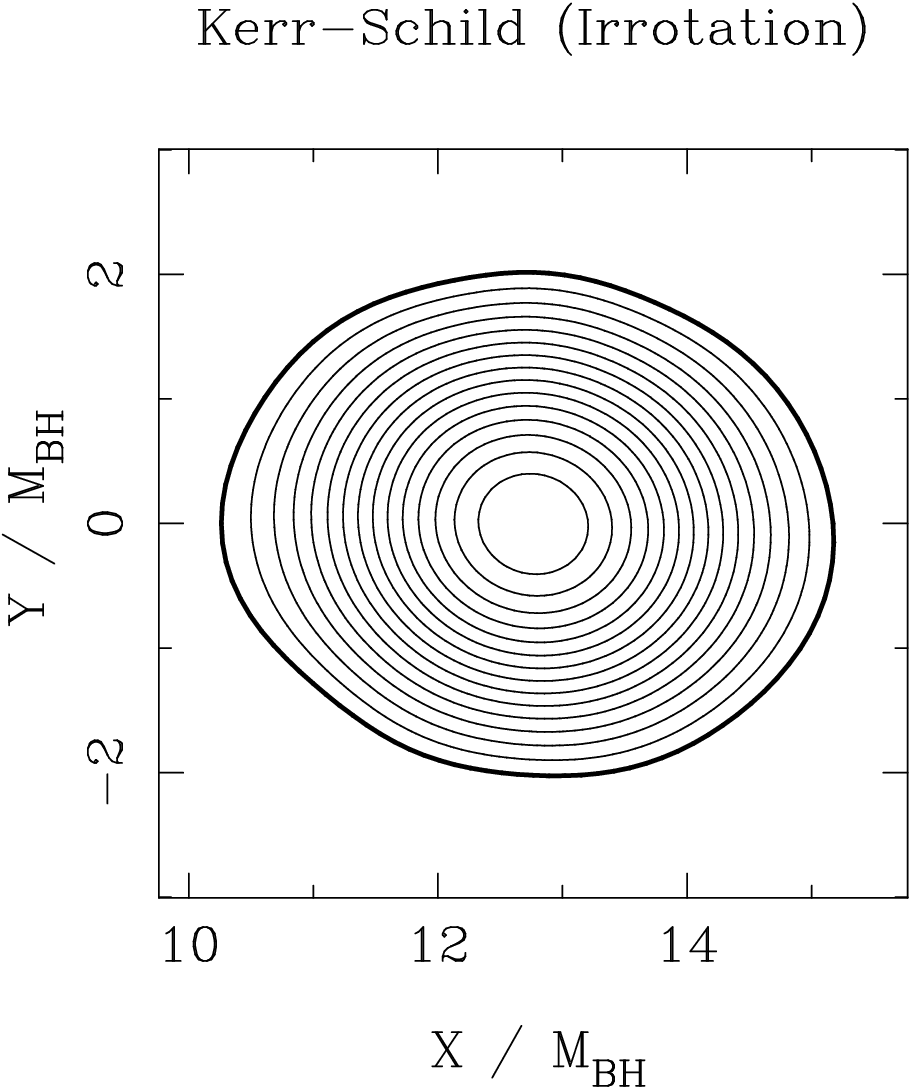}
  \hspace{10pt}
  \includegraphics[height=7cm]{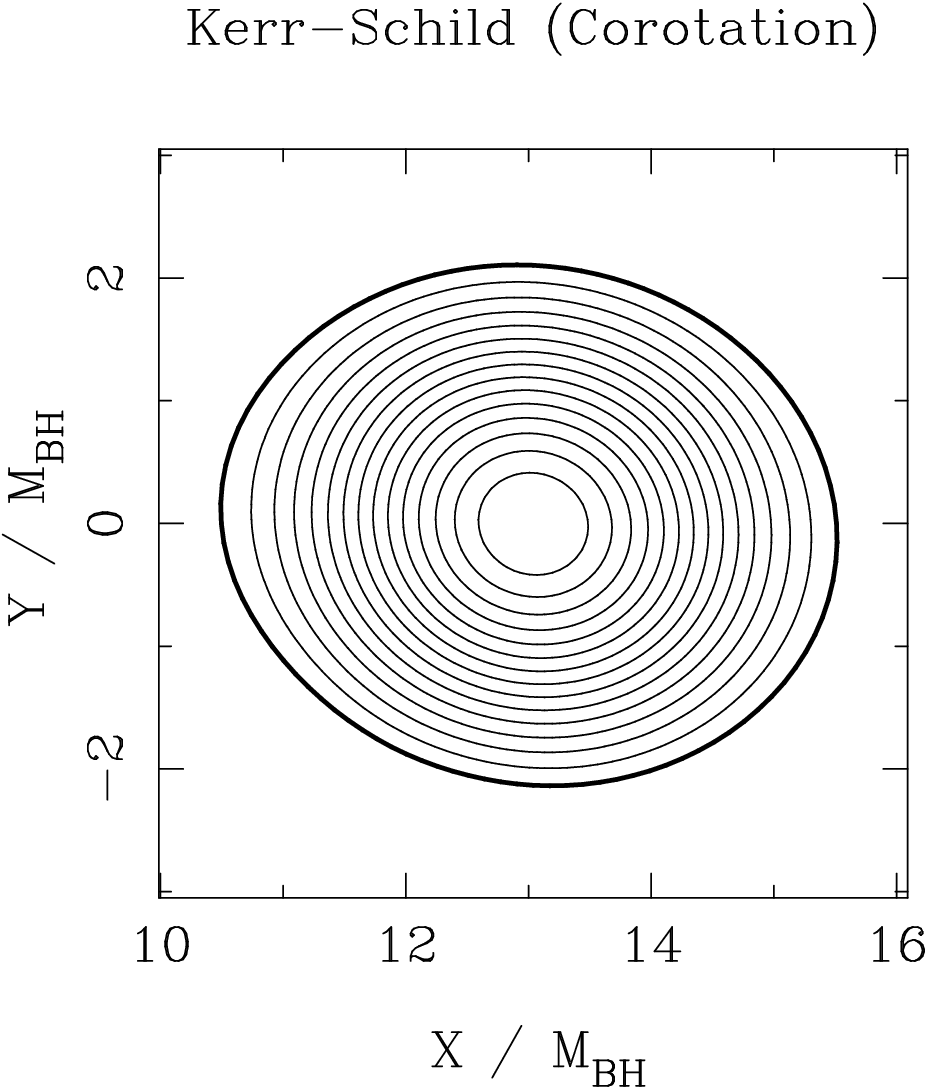}
\end{center}
\caption{Baryon rest-mass density contours for a neutron star with
$\bar{M}_{\rm NS}=0.05$ in the Kerr-Schild background, at the point of
closest binary separation along our sequence.  
The left and right panels show the irrotational and corotating cases,
respectively.
The position of the density maximum is
$X_{\rm NS}/M_{\rm BH}=12.8$ for the irrotational case
and $X_{\rm NS}/M_{\rm BH}=13.0$ for the corotating one.}
\label{fig:dens_KS005}
\end{figure*}

\begin{figure*}[ht]
\begin{center}
  \includegraphics[height=7cm]{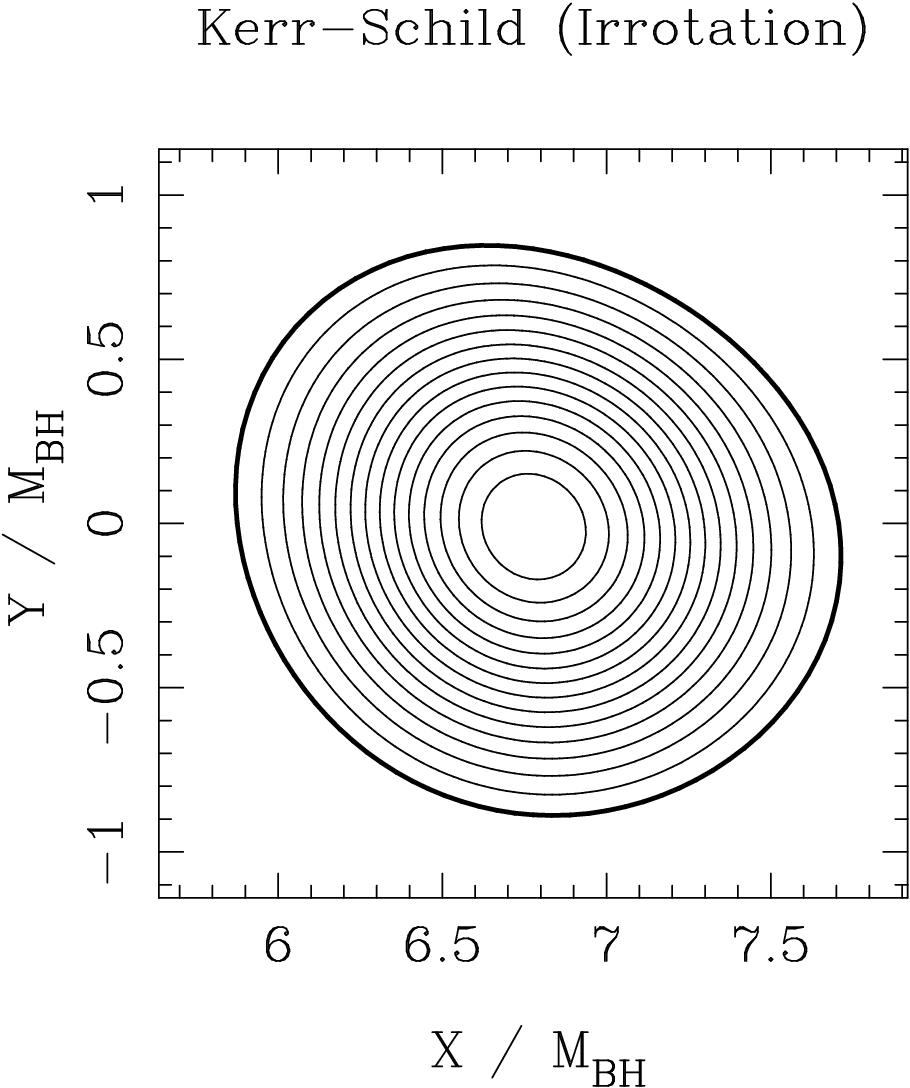}
  \hspace{10pt}
  \includegraphics[height=7cm]{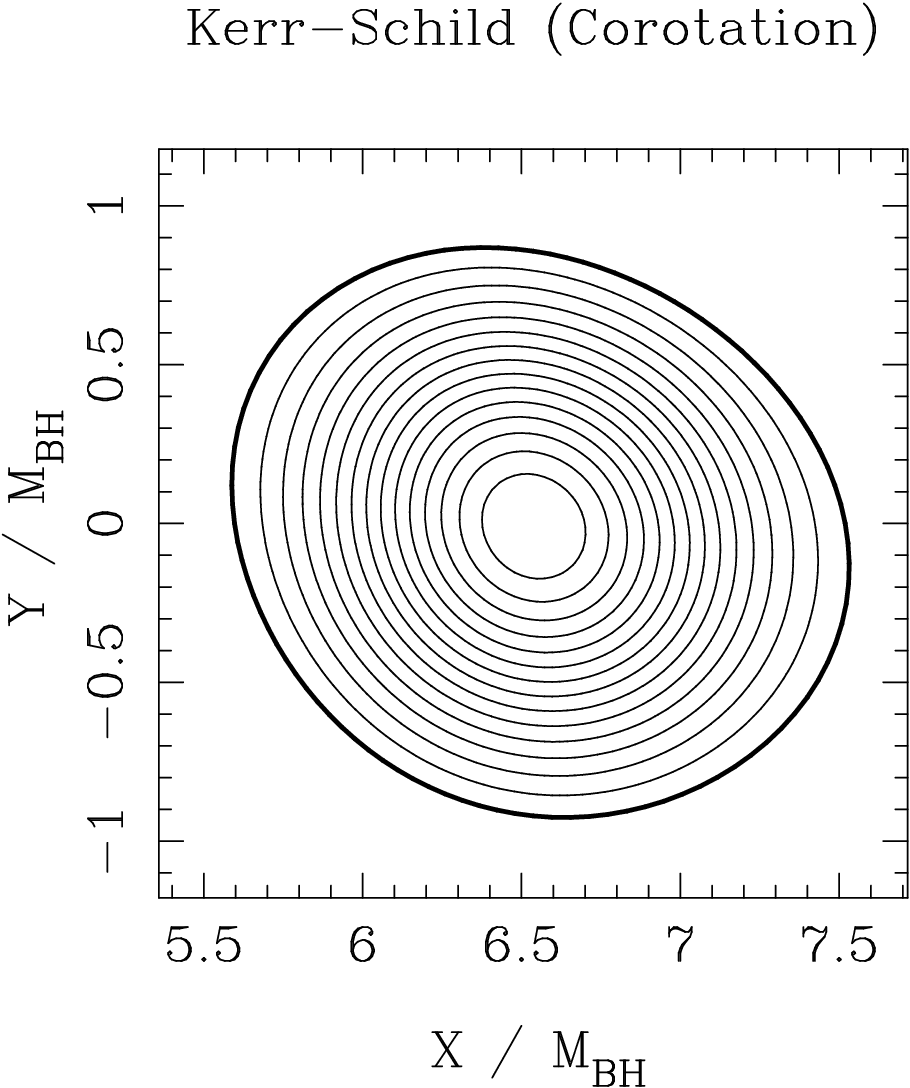}
\end{center}
\caption{
Same as Fig.~\ref{fig:dens_KS005}, but for $\bar{M}_{\rm NS}=0.1$.
The position of the density maximum is
$X_{\rm NS}/M_{\rm BH}=6.78$ for the irrotational case
and $X_{\rm NS}/M_{\rm BH}=6.54$ for the corotating one.}
\label{fig:dens_KS01}
\end{figure*}

\begin{figure*}[ht]
\begin{center}
  \includegraphics[height=7cm]{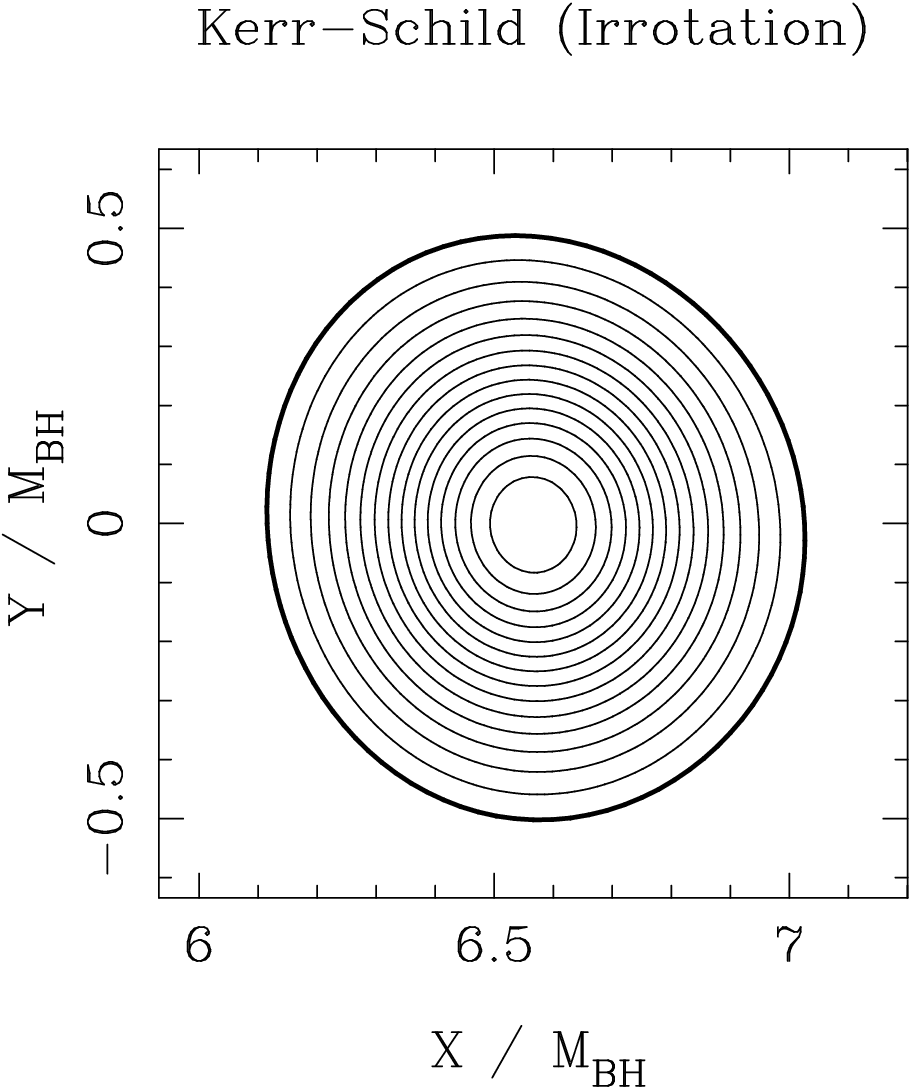}
  \hspace{10pt}
  \includegraphics[height=7cm]{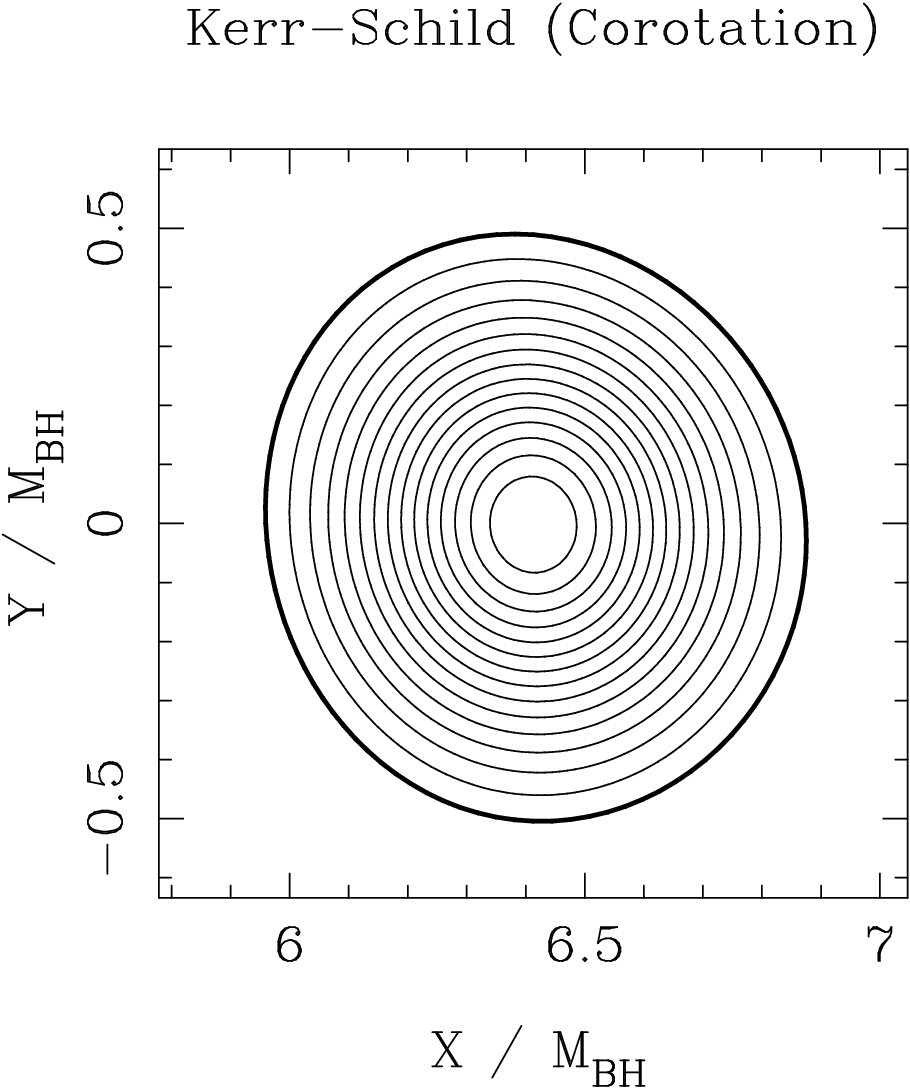}
\end{center}
\caption{
Same as Fig.~\ref{fig:dens_KS005}, but for $\bar{M}_{\rm NS}=0.15$.
The position of the density maximum is
$X_{\rm NS}/M_{\rm BH}=6.57$ for the irrotational case
and $X_{\rm NS}/M_{\rm BH}=6.41$ for the corotating one.}
\label{fig:dens_KS015}
\end{figure*}

In Figs.~\ref{fig:dens_KS005} -- \ref{fig:dens_KS015}, we show contours
of the neutron star baryon rest-mass density profile at the closest
separation we calculate for neutron star masses
$\bar{M}_{\rm NS}=0.05$, 0.1, and 0.15, respectively.
In each figure, the left panel shows the irrotational case and the right
panel the corotating one.
These figures correspond to the final lines in
Tables \ref{table:seq_K005} -- \ref{table:seq_K015}.
Since the total shift vector induces an asymmetry in the neutron star
with respect to the $X\!\!-\!\!Z$ plane for a Kerr-Schild background,
one can see a visible tilt in the stellar figure.

In the left panel of Fig. \ref{fig:dens_KS005},
one can see a slight oscillation on the stellar surface.
During the computation, we fit the boundary of the innermost domain
to the stellar surface, allowing us to solve equilibrium configurations
accurately, particularly in irrotational cases.
Once a cusp develops on the stellar surface,
it becomes impossible to adapt the innermost domain to the stellar surface.
Using spectral methods, we express all quantities by
summation over a finite number of differentiable functions.
If we apply this method to such cusp-like figures,
large numerical errors are induced.
This situation is known as the Gibbs phenomenon.
Since the stellar surface becomes highly distorted
for very close configurations,
even prior to the appearance of a cusp,
we have to stop the sequence when such oscillations appear.

In order to investigate how close our results come to the proper tidal
disruption points, we introduce a sensitive numerical indicator for the
mass-shedding point, defined as
\begin{equation}\label{eq:chi}
  \chi \equiv {(\partial H_{\rm ent}/\partial r)_{\rm eq, comp} \over
    (\partial H_{\rm ent}/\partial r)_{\rm pole}},
\end{equation}
where the numerator denotes the radial derivative of enthalpy of the
neutron star toward the companion star in the equatorial plane on the
$X$-axis and the denominator that toward the pole of the neutron star.
Both terms are evaluated on the stellar surface.
This indicator takes the value unity at infinite orbital separation,
and goes to zero at the mass-shedding point.

\begin{figure*}[ht]
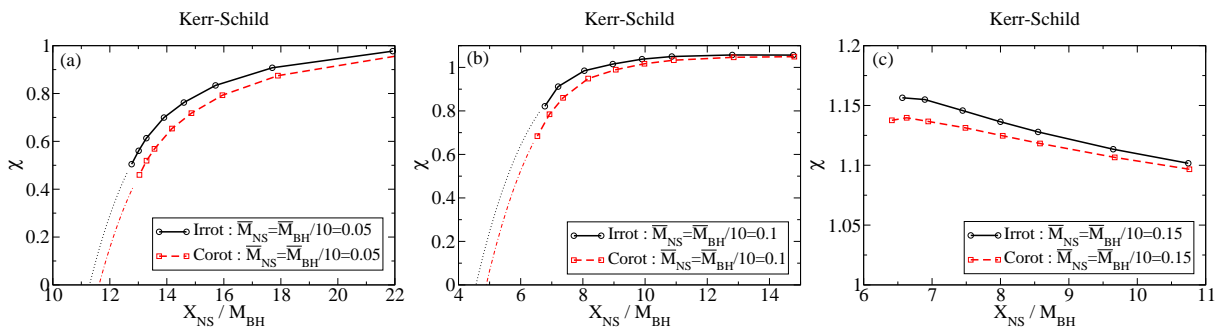

\vspace{0.5cm}
\begin{center}
  \includegraphics[height=4.2cm]{fig8a_chi_K005.eps}
  \includegraphics[height=4.2cm]{fig8b_chi_K01.eps}
  \includegraphics[height=4.2cm]{fig8c_chi_K015.eps}
\end{center}
\caption{Mass-shedding indicator $\chi$, defined by
Eq.~(\protect\ref{eq:chi}), as a function of the orbital
separation for the neutron star mass (a)$\bar{M}_{\rm NS}=0.05$,
(b)$\bar{M}_{\rm NS}=0.1$ and (c)$\bar{M}_{\rm NS}=0.15$,
in Kerr-Schild backgrounds.
Mass-shedding will begin when $\chi=0$.
The thick solid line with circles represents the irrotational case
and the thick dashed line with squares the corotating one.
The thin dotted and dot-dashed lines are extrapolations
for the results of the irrotational and corotating cases, respectively.}
\label{fig:chi_KS}
\end{figure*}

We show numerical results for $\chi$ in Fig.~\ref{fig:chi_KS}.
The solid line with circles represents the irrotational case and
the dashed line with squares the corotating one.
As explained above, it is difficult for the spectral method to treat
cusp-like figures, so we stop the calculation of sequences before
reaching $\chi=0$, which would correspond to the mass-shedding point.
However, we can extrapolate our results in polynomial functions
to predict the orbital separation at the tidal disruption point, $\chi=0$.
The extrapolations are shown as a dotted line for the irrotational
case and a dot-dashed line for the corotating case
in Figs. \ref{fig:chi_KS}(a) and (b).
For a neutron star of mass $\bar{M}_{\rm NS}=0.15$, we do not attempt
to extrapolate our results to the tidal disruption limit
because the closest separations we can calculate are
too far away, for both rotation states, to make a reasonable fit.
Clearly the extrapolation introduces considerable error,
so that we are able to predict the tidal separation only
with modest accuracy.

We find that the state of rotation has a small effect of a few percent
on the tidal separation, which is in agreement with earlier findings
(e.g. \cite{WiggL00}).
For $\bar{M}_{\rm NS}=0.05$ we find
approximately $r_{\rm tid} = 11.5 M_{\rm BH}$
(which is in good agreement with the value $11.9 M_{\rm BH}$ found
in BSS), and for $\bar{M}_{\rm NS}=0.1$ we find approximately
$r_{\rm tid} = 5 M_{\rm BH}$.

Finally, we examine the relative change in the central energy density
of the neutron star to that of an isolated, spherical neutron star with
the same baryon rest-mass.
Here, the total energy density
and the relative change of its central value are, respectively, defined by
$e \equiv \rho_0 +\rho_i$ and
\begin{equation}
  \delta e \equiv {e_c \over e_{c,0}} - 1,
\end{equation}
where $e_c$ is the central energy density of the neutron star and
$e_{c,0}$ that of a spherical neutron star
with the same baryon rest-mass.
We see from Tables \ref{table:seq_K005} -- \ref{table:seq_K015}
that the total energy density decreases as the orbital separation
decreases for all sequences.
At fixed orbital separation and neutron star mass, the decrease in the
total energy density is larger for the corotating case than for the
irrotational one.
We attribute this effect to the neutron star spin, since the rotation
expands the star's volume and results in a larger decrease in the central
energy density.
Roughly speaking, the amount of relative decrease of the central energy
density at the mass-shedding point is $\sim 5\%$ for an irrotational
neutron star of mass $\bar{M}_{\rm NS}=0.05$,
and $\sim 10\%$ for a corotating one,
if we extrapolate our results.
For a neutron star of mass $\bar{M}_{\rm NS}=0.1$, the relative decrease is
about $\sim 15\%$ for the irrotational case and about $\sim 20\%$ for the
corotating one.
This implies that a more massive neutron star undergoes a larger
decrease in the energy density than a less massive one.
We note that the central energy density decreases monotonically,
even when the half-diameter on the $X$-axis, $\bar{r}_e$, decreases,
as we find for sequences with $\bar{M}_{\rm NS}=0.1$ and 0.15.
This occurs, in part, because $\bar{r}_e$ is measured as
a coordinate length, not as the proper length.
Also, the diameter along the $X$-axis is not the
primary axis in Kerr-Schild coordinates,
because the stellar configuration is tilted due to asymmetries.

\subsection{The isotropic background}

\begin{table*}[ht]
\caption{Same as Table \ref{table:seq_K005}, but for isotropic backgrounds.}
\begin{center}
\begin{tabular}{rccccccc} \hline\hline
  \multicolumn{8}{c}{Isotropic: $\bar{M}_{\rm NS}=0.05$,
    $M_{\rm NS}^G/r_0=0.0415$, $\bar{r}_0=1.1289$} \\ \hline\hline
  \multicolumn{8}{c}{Irrotation} \\
  $X_{\rm NS}/r_e$&$X_{\rm NS}/M_{\rm BH}$&$\Omega M_{\rm BH}$&
  $~~~\bar{r}_e~~~$&$~~q_{\rm max}~~$&$\delta e$&$~~~~\chi~~~~$&
  $R_{\rm B}/r_0$ \\ \hline
  10.0&21.9 &9.11(-3) &1.10 &0.0235 &-1.78(-3) &0.942 &8 \\
  8.0 &17.6 &1.24(-2) &1.10 &0.0235 &-3.52(-3) &0.883 &7 \\
  6.0 &13.7 &1.78(-2) &1.14 &0.0233 &-1.01(-2) &0.719 &5 \\
  5.0 &12.0 &2.13(-2) &1.20 &0.0231 &-2.08(-2) &0.514 &4.5 \\
  4.7 &11.6 &2.24(-2) &1.23 &0.0230 &-2.50(-2) &0.406 &4 \\ \hline\hline
  \multicolumn{8}{c}{Corotation} \\
  $X_{\rm NS}/r_e$&$X_{\rm NS}/M_{\rm BH}$&$\Omega M_{\rm BH}$&$\bar{r}_e$&
  $q_{\rm max}$&$\delta e$&$\chi$&$R_{\rm B}/r_0$ \\ \hline
  10.0&22.0 &9.04(-3) &1.10 &0.0234 &-7.49(-3) &0.925 &8 \\
  8.0 &17.8 &1.23(-2) &1.11 &0.0232 &-1.42(-2) &0.855 &7 \\
  6.0 &13.9 &1.74(-2) &1.16 &0.0228 &-3.25(-2) &0.681 &5 \\
  5.0 &12.2 &2.07(-2) &1.22 &0.0223 &-5.35(-2) &0.495 &4.5 \\
  4.7 &11.8 &2.17(-2) &1.26 &0.0221 &-6.21(-2) &0.413 &4 \\ \hline
\end{tabular}
\end{center}
\label{table:seq_C005}
\end{table*}

\begin{table*}[ht]
\caption{Same as Table \ref{table:seq_C005}, but for $\bar{M}_{\rm NS}=0.1$}
\begin{center}
\begin{tabular}{rccccccc} \hline\hline
  \multicolumn{8}{c}{Isotropic: $\bar{M}_{\rm NS}=0.1$,
    $M_{\rm NS}^G/r_0=0.0879$, $\bar{r}_0=0.98972$} \\ \hline\hline
  \multicolumn{8}{c}{Irrotation} \\
  $X_{\rm NS}/r_e$&$X_{\rm NS}/M_{\rm BH}$&$\Omega M_{\rm BH}$&
  $~~~\bar{r}_e~~~$&$~~q_{\rm max}~~$&$\delta e$&$~~~~\chi~~~~$&
  $R_{\rm B}/r_0$ \\ \hline
  14.0&12.9 &1.94(-2) &0.919 &0.0586 &-3.62(-3) &0.980 &8 \\
  10.0&9.01 &3.15(-2) &0.901 &0.0583 &-8.74(-3) &0.936 &7 \\
  8.0 &7.16 &4.27(-2) &0.895 &0.0579 &-1.66(-2) &0.863 &6 \\
  6.0 &5.51 &5.96(-2) &0.919 &0.0568 &-3.70(-2) &0.659 &4.5 \\
  5.6 &5.24 &6.35(-2) &0.936 &0.0563 &-4.55(-2) &0.577 &4.5 \\ \hline\hline
  \multicolumn{8}{c}{Corotation} \\
  $X_{\rm NS}/r_e$&$X_{\rm NS}/M_{\rm BH}$&$\Omega M_{\rm BH}$&$\bar{r}_e$&
  $q_{\rm max}$&$\delta e$&$\chi$&$R_{\rm B}/r_0$ \\ \hline
  14.0&12.9 &1.93(-2) &0.921 &0.0585 &-6.62(-3) &0.973 &8 \\
  10.0&9.07 &3.12(-2) &0.907 &0.0579 &-1.67(-2) &0.918 &7 \\
  8.0 &7.25 &4.20(-2) &0.907 &0.0571 &-3.11(-2) &0.837 &6 \\
  6.0 &5.65 &5.78(-2) &0.942 &0.0550 &-6.79(-2) &0.634 &5 \\
  5.6 &5.37 &6.16(-2) &0.958 &0.0545 &-7.81(-2) &0.564 &4.5 \\ \hline
\end{tabular}
\end{center}
\label{table:seq_C01}
\end{table*}

\begin{table*}[ht]
\caption{Same as Table \ref{table:seq_C005}, but for $\bar{M}_{\rm NS}=0.15$}
\begin{center}
\begin{tabular}{rccccccc} \hline\hline
  \multicolumn{8}{c}{Isotropic: $\bar{M}_{\rm NS}=0.15$,
    $M_{\rm NS}^G/r_0=0.145$, $\bar{r}_0=0.81526$} \\ \hline\hline
  \multicolumn{8}{c}{Irrotation} \\
  $X_{\rm NS}/r_e$&$X_{\rm NS}/M_{\rm BH}$&$\Omega M_{\rm BH}$&
  $~~~\bar{r}_e~~~$&$~~q_{\rm max}~~$&$\delta e$&$~~~~\chi~~~~$&
  $R_{\rm B}/r_0$ \\ \hline
  20.0&9.75 &2.83(-2) &0.732 &0.126 &-5.24(-3) &0.995 &8 \\
  16.0&7.59 &3.96(-2) &0.712 &0.126 &-9.67(-3) &0.988 &8 \\
  12.0&5.46 &6.06(-2) &0.682 &0.124 &-2.31(-2) &0.959 &7 \\
  10.0&4.43 &7.82(-2) &0.665 &0.122 &-4.22(-2) &0.913 &6 \\
  9.0 &3.96 &8.94(-2) &0.659 &0.120 &-6.06(-2) &0.868 &5.5 \\ \hline\hline
  \multicolumn{8}{c}{Corotation} \\
  $X_{\rm NS}/r_e$&$X_{\rm NS}/M_{\rm BH}$&$\Omega M_{\rm BH}$&$\bar{r}_e$&
  $q_{\rm max}$&$\delta e$&$\chi$&$R_{\rm B}/r_0$ \\ \hline
  20.0&9.77 &2.82(-2) &0.733 &0.126 &-7.23(-3) &0.992 &8 \\
  16.0&7.61 &3.94(-2) &0.714 &0.125 &-1.35(-2) &0.982 &8 \\
  12.0&5.50 &6.00(-2) &0.687 &0.123 &-3.18(-2) &0.948 &7 \\
  10.0&4.49 &7.68(-2) &0.674 &0.120 &-5.60(-2) &0.899 &6 \\
  9.4 &4.20 &8.30(-2) &0.671 &0.119 &-6.74(-2) &0.874 &5.5 \\ \hline
\end{tabular}
\end{center}
\label{table:seq_C015}
\end{table*}

\begin{figure*}[ht]
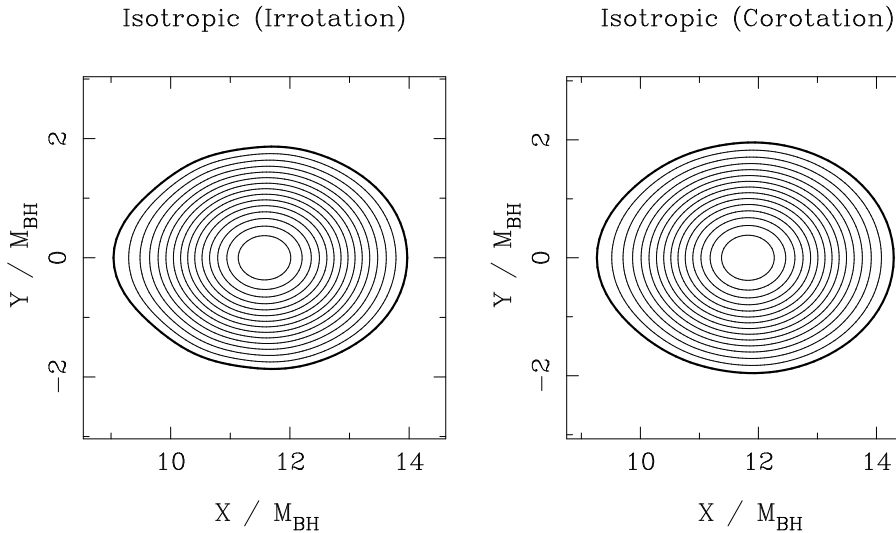

\begin{center}
  \includegraphics[height=7cm]{fig9a_dens_xy_C005i.eps}
  \hspace{10pt}
  \includegraphics[height=7cm]{fig9b_dens_xy_C005c.eps}
\end{center}
\caption{Baryon rest-mass density contours for neutron stars
with $\bar{M}_{\rm NS}=0.05$ in an isotropic background, at the
point of closest binary separation along our sequence.  
The left and right panels show the irrotational and corotating cases,
respectively.
The position of the density maximum is
$X_{\rm NS}/M_{\rm BH}=11.6$ for the irrotational case
and $X_{\rm NS}/M_{\rm BH}=11.8$ for the corotating one.}
\label{fig:dens_CF005}
\end{figure*}

\begin{figure*}[ht]
\begin{center}
  \includegraphics[height=7cm]{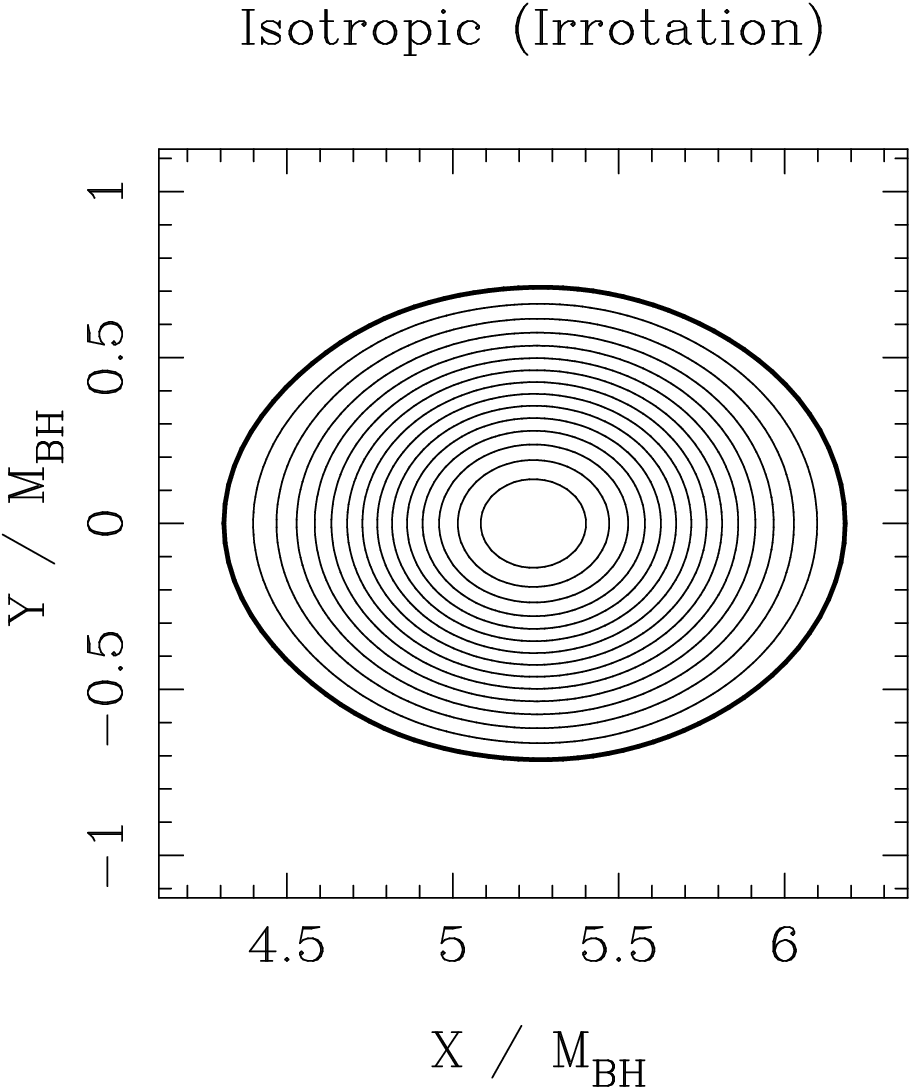}
  \hspace{10pt}
  \includegraphics[height=7cm]{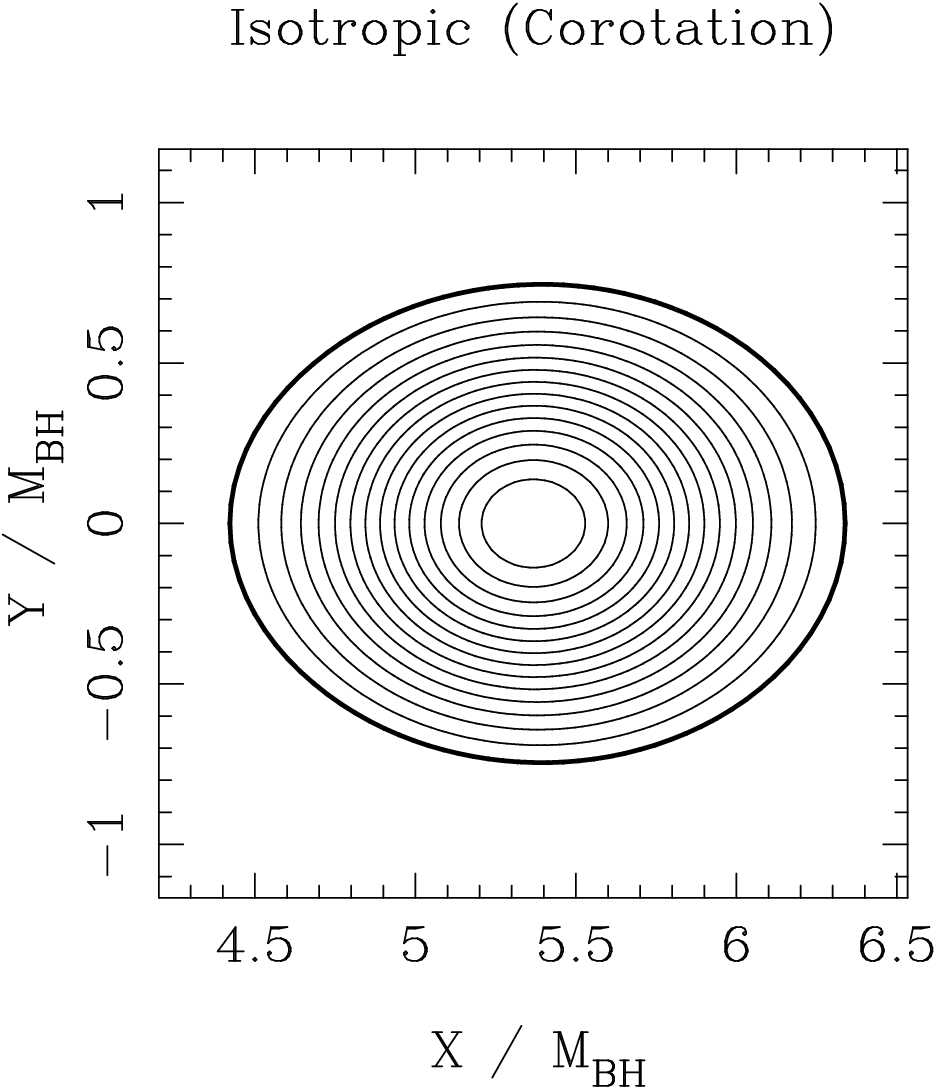}
\end{center}
\caption{Same as Fig.~\ref{fig:dens_CF005}, but for $\bar{M}_{\rm NS}=0.1$.
The position of the density maximum is
$X_{\rm NS}/M_{\rm BH}=5.24$ for the irrotational case
and $X_{\rm NS}/M_{\rm BH}=5.37$ for the corotating one.}
\label{fig:dens_CF01}
\end{figure*}

\begin{figure*}[ht]
\begin{center}
  \includegraphics[height=7cm]{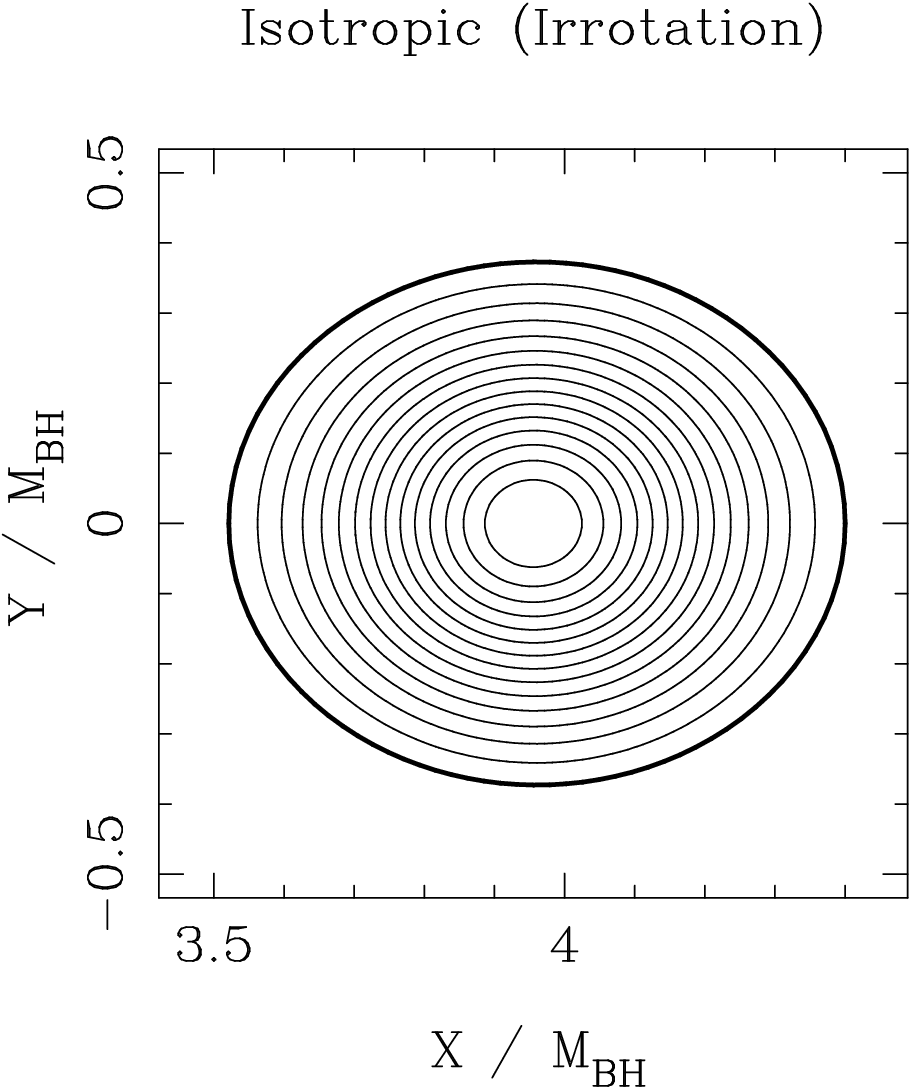}
  \hspace{10pt}
  \includegraphics[height=7cm]{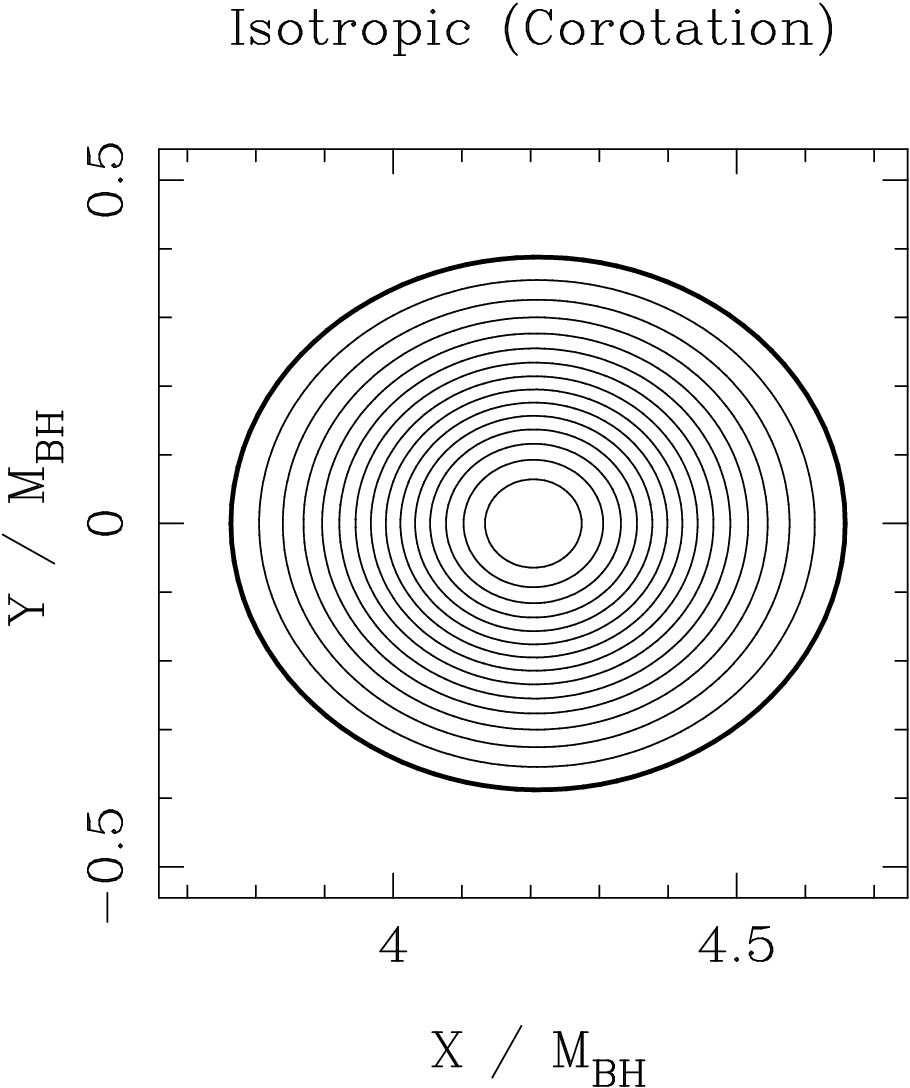}
\end{center}
\caption{Same as Fig.~\ref{fig:dens_CF005}, but for $\bar{M}_{\rm NS}=0.15$.
The position of the density maximum is
$X_{\rm NS}/M_{\rm BH}=3.96$ for the irrotational case and
$X_{\rm NS}/M_{\rm BH}=4.20$ for the corotating one.}
\label{fig:dens_CF015}
\end{figure*}

In Figs. \ref{fig:dens_CF005} -- \ref{fig:dens_CF015},
we show contours of the baryon rest-mass density for neutron stars
in an isotropic background. 
As in the Kerr-Schild background, the left panel is for the
irrotational case and the right panel for the corotating one.
These figures are taken from the configurations shown on the 
last lines of Tables \ref{table:seq_C005} -- \ref{table:seq_C015},
the closest separation we compute for each sequence.
One can clearly see that there exists a symmetry with respect to
the $X\!\!-\!\!Z$ plane in addition to the $X\!\!-\!\!Y$ plane symmetry
(equatorial plane symmetry) in isotropic background.
The half-diameter of the neutron star along the $X$-axis, $\bar{r}_e$,
should be elongated by the tidal force from the black hole, 
but as in the Kerr-Schild background it decreases as the orbital separation
decreases for $\bar{M}_{\rm NS}=0.1$ and 0.15, and then increases
at closer separations for $\bar{M}_{\rm NS}=0.1$
because of coordinate effects.

\begin{figure*}[ht]
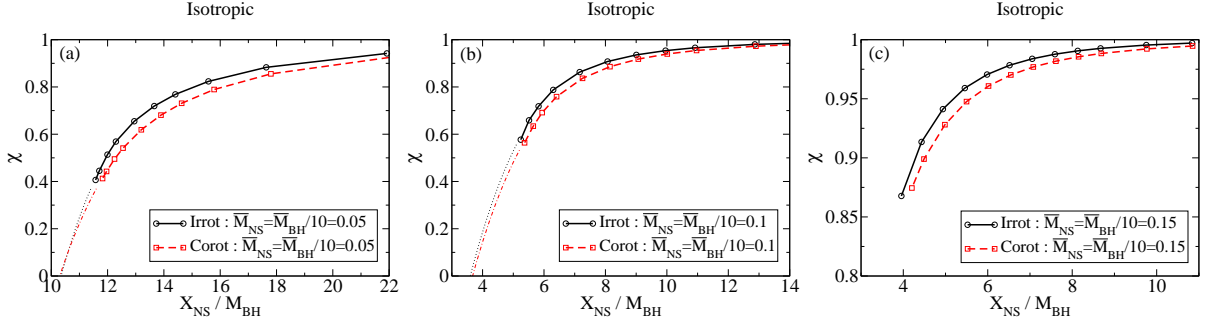

\vspace{0.5cm}
\begin{center}
  \includegraphics[height=4.2cm]{fig12a_chi_C005.eps}
  \includegraphics[height=4.2cm]{fig12b_chi_C01.eps}
  \includegraphics[height=4.2cm]{fig12c_chi_C015.eps}
\end{center}
\caption{Mass-shedding indicator $\chi$, defined by
Eq.~(\protect\ref{eq:chi}), as a function of the orbital
separation for neutron star of mass (a)$\bar{M}_{\rm NS}=0.05$,
(b)$\bar{M}_{\rm NS}=0.1$ and (c)$\bar{M}_{\rm NS}=0.15$
in isotropic backgrounds.
The thick solid line with circles represents the irrotational case
and the thick dashed line with squares the corotating one.
The thin dotted and dot-dashed lines are extrapolations
of the results for the irrotational and corotating cases, respectively.}
\label{fig:chi_CF}
\end{figure*}

The indicator for mass-shedding from the neutron star, $\chi$,
is shown in Fig. \ref{fig:chi_CF}.
Solid lines with circles and dashed lines with squares in each panel
represent the irrotational case and corotating one, respectively.
The dotted lines are extrapolations of our data for the irrotational case
and dot-dashed lines those for the corotating case shown
in Figs. \ref{fig:chi_CF} (a) and (b).
For isotropic backgrounds, $\chi$ does not exceed unity even
for high compactness,
and decreases monotonically when the orbital separation decreases.
From the extrapolations, we see that the orbital separations at the
mass-shedding point are $X_{\rm NS} \sim 10.5 M_{\rm BH}$ for
$\bar{M}_{\rm NS}=0.05$ and $X_{\rm NS} \sim 4 M_{\rm BH}$ for
$\bar{M}_{\rm NS}=0.1$ with an error of approximately $10\%$.
These values are coordinate separations, and should therefore not be
compared immediately with the results in a Kerr-Schild background.
We will compare our findings in much more detail in
Section \ref{sec:comparison}.
As before we do not extrapolate for $\bar{M}_{\rm NS}=0.15$
because the closest separation we can calculate is still far from the
mass-shedding point.

\begin{figure*}[ht]
\vspace{0.5cm}
\begin{center}
  \includegraphics[height=6.5cm]{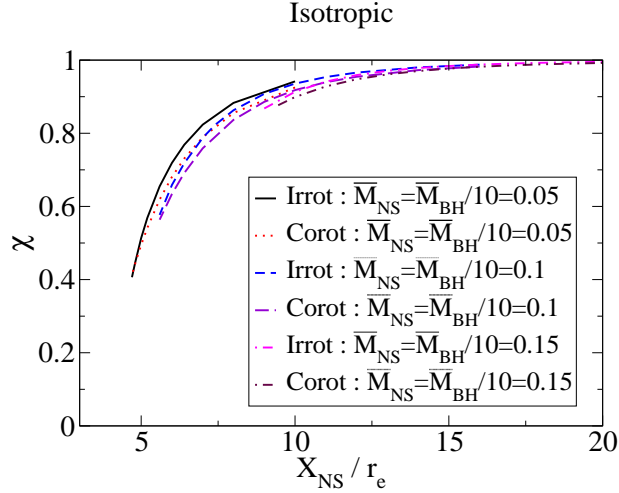}
\end{center}
\caption{Mass-shedding indicator $\chi$ as a function of the
orbital separation $X_{\rm NS}/r_e$.
Solid, dashed, and dot-dashed lines denote the irrotational case
with neutron star masses $\bar{M}_{\rm NS}=0.05$, 0.1, and 0.15,
respectively.
Dotted, long-dashed, and dot-dot-dashed lines are for the corotating case
with $\bar{M}_{\rm NS}=0.05$, 0.1, and 0.15.}
\label{fig:chi-X_CF}
\end{figure*}

Since the indicator $\chi$ does not exceed unity for any sequence in
an isotropic background and decreases monotonically toward zero,
it is convenient to compare sequences for different neutron star masses
$\bar{M}_{\rm NS}$ in the same figure.
In Fig. \ref{fig:chi-X_CF}, we show the mass-shedding indicator, $\chi$,
as a function of $X_{\rm NS}/r_e$, for several different sequences.
Interestingly, we see that $\chi$ is almost independent of the neutron
star mass or spin, determined almost entirely by the normalized separation
$X_{\rm NS}/r_e$.
A more massive neutron star has slightly smaller value of $\chi$, indicating
that it would be disrupted at a larger normalized separation
$X_{\rm NS}/r_e$.
We can see the same behavior in the binary neutron stars,
for example, in Fig. 7 of \cite{TanigG02b}.

The central energy density again decreases for all sequences with the
binary separation, as we found for all sequences in a Kerr-Schild
background.

\section{Comparison between Kerr-Schild and isotropic backgrounds}
\label{sec:comparison}

Kerr-Schild coordinates and isotropic coordinates represent a
Schwarzschild black hole in two different coordinate systems that
differ not only in the spatial coordinates but also in the
time-coordinate.
Spatial slices in the two coordinate systems therefore represent distinct
slices of Schwarzschild, and there is no reason to expect that solving
the conformal thin-sandwich equations on these different slices will
result in physically equivalent solutions.

It is of interest, then, to compare our results for Kerr-Schild and
isotropic backgrounds.
Most interesting, of course, are comparisons of coordinate-independent
quantities, in particular the angular velocity
(for example at tidal break-up), and the maximum density.
We also include comparisons of the effective enthalpy and the shift vector,
which behaves very differently in the two coordinate systems.

\subsection{Orbital angular velocity at the mass-shedding point}

\begin{figure*}[ht]
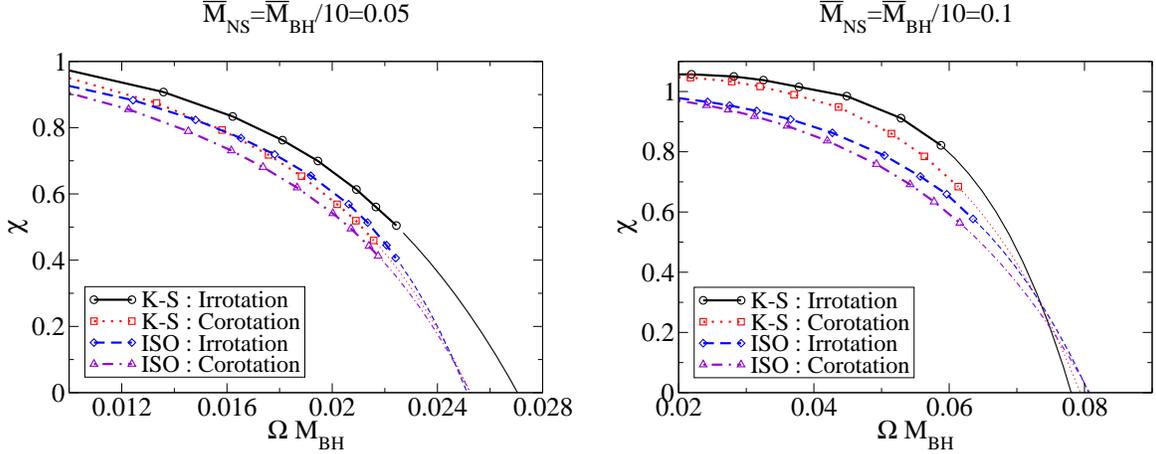

\vspace{0.5cm}
\begin{center}
  \includegraphics[height=6cm]{fig14a_omega_005.eps}
  \hspace{10pt}
  \includegraphics[height=6cm]{fig14b_omega_01.eps}
\end{center}
\caption{The mass-shedding indicator $\chi$
as a function of the orbital angular velocity.
The left panel shows sequences with neutron star mass
$\bar{M}_{\rm NS}=0.05$, the right $\bar{M}_{\rm NS}=0.1$.
K-S and ISO denote sequences performed in the respective backgrounds.}
\label{fig:omega}
\end{figure*}

As we explained previously, we stop the computation of each sequence
before the mass-shedding point since it is impossible for the spectral
method to treat cusp-like figures.
Thus, we have to extrapolate all sequences to the mass-shedding point
to estimate the value of the orbital angular velocity there.
In Fig.~\ref{fig:omega}, we show the mass-shedding indicator $\chi$
as a function of the orbital angular velocity.
The indicator $\chi$ decreases rapidly as the separation decreases,
or equivalently as the orbital angular velocity increases,
so that extrapolating the curves using only our calculated values at
large and medium separations may give predictions with large errors,
especially for the irrotational case in the Kerr-Schild background
with $\bar{M}_{\rm NS}=0.05$;
in cases where the sequences continue to smaller separations we expect
an error of no more than $\sim 10\%$ in our measure of the terminal
angular velocity.

For both Kerr-Schild and isotropic backgrounds, we find a value
$\Omega M_{\rm BH} \sim 0.025$ for the corotating case with neutron star
mass $\bar{M}_{\rm NS}=0.05$, in good agreement with the results of
BSS, who find $\Omega M_{\rm BH}=0.0241$ (see their Table 1).

Finally, we comment on the discrepancy in the extrapolation curve
for the irrotational case in the Kerr-Schild background
with $\bar{M}_{\rm NS}=0.05$.
In this case, since we do not have results in the range
$0 < \chi < 0.5$, we cannot draw an accurate extrapolation curve,
as mentioned above, and thus cannot use it to predict
the orbital angular velocity at the mass-shedding point with a high
degree of certainty.
However, we can predict that the real orbital angular velocity at the
mass-shedding point should be smaller than $\Omega M_{\rm BH} \sim 0.027$
the value given by Fig.~\ref{fig:omega},
because the parameter $\chi$ decreases more rapidly at larger 
orbital angular velocities for the irrotational case
as compared to the corotating one for the range $\chi < 0.5$
(see Section V.B of \cite{TanigGB01}).
This means that the data points for the irrotational case
of the Kerr-Schild background with $\bar{M}_{\rm NS}=0.05$
are most likely to lie below the extrapolated line (solid line)
in reality.

\subsection{Maximum value of the density quantity}

\begin{figure*}[ht]
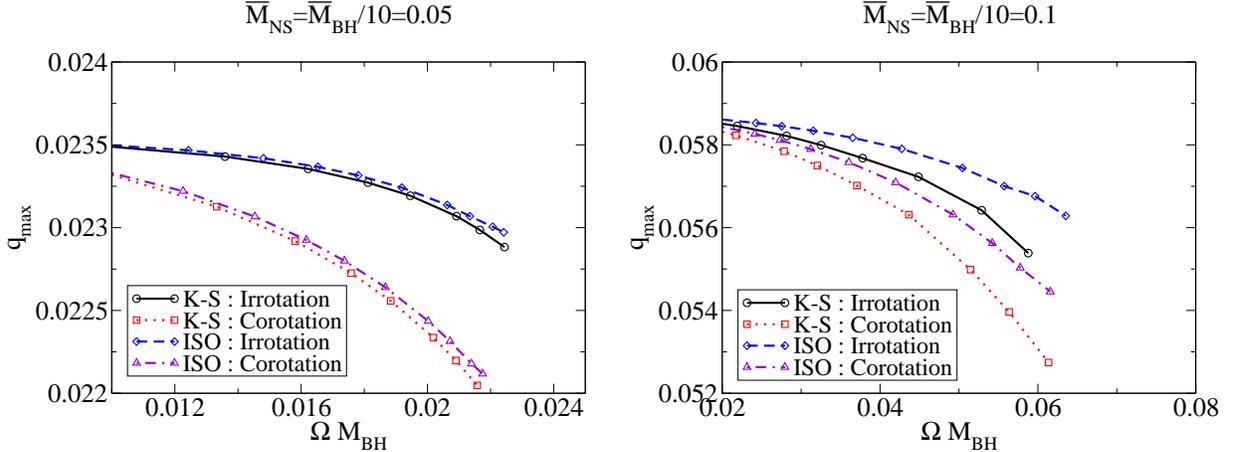

\vspace{0.5cm}
\begin{center}
  \includegraphics[height=6cm]{fig15a_qmax_005.eps}
  \hspace{10pt}
  \includegraphics[height=6cm]{fig15b_qmax_01.eps}
\end{center}
\caption{Maximum value of the density quantity $q_{\rm max}$
as a function of the orbital angular velocity.
The left panel corresponds to a neutron star of mass
$\bar{M}_{\rm NS}=0.05$ the right panel $\bar{M}_{\rm NS}=0.1$.
K-S and ISO denote the respective backgrounds.
}
\label{fig:qmax}
\end{figure*}

The density parameter $q$ is a measure of the density as seen by a
comoving observer, and is therefore gauge-invariant.
In Fig.~\ref{fig:qmax} we show its maximum value in the star as a
function of the orbital angular velocity, which is also
coordinate-independent.
As we have discussed before, the maximum density decreases with binary
separation in all cases, and decreases more rapidly for corotating
configurations.

For $\bar{M}_{\rm NS}=0.05$ we find only very small differences
between Kerr-Schild and isotropic backgrounds (less than 1\%).
However, for $\bar{M}_{\rm NS}=0.1$ the compaction parameter
$M_{\rm NS}^G/r_0$ ($M_{\rm NS}^G$ is the gravitational mass of an
isolated neutron star and $r_0$ its radius) and
hence relativistic effects are significantly larger,
the differences increase noticeably.
The difference between the two backgrounds is now in the order of 5\%,
meaning that the effect of the background is almost as large as that
of the presence of a binary companion.
We find in all cases the Kerr-Schild background leads to a greater
decrease in the central density than an isotropic background.

\subsection{Effective enthalpy field}

For corotating configurations the definition of the enthalpy can be
extended to regions outside the neutron star by virtue of
Eq.~(\ref{eq:iEuler}).
As discussed in BSS this ``effective enthalpy field'' plays
the role of an effective potential, allows for the definition of a Roche
lobe in a fully relativistic context, and is very useful for locating
the onset of tidal disruption.
To see this, we evaluate (\ref{eq:iEuler}) for corotating binaries
\begin{equation}
  H_{\rm ent} = C -\nu +\ln \gamma_0,
\end{equation}
where $C$ is a constant.
Inserting the definition of $\nu$ (Eq. (\ref{eq:nu})) and $\gamma_0$
(Eq. (\ref{eq:gamma0})) for the corotating case, we obtain
\begin{equation}
  H_{\rm ent} = C -\ln \alpha -{1 \over 2}
  \ln \Bigl( 1 -{\gamma_{ij} \beta^i \beta^j \over \alpha^2} \Bigr).
  \label{eq:effect_ent}
\end{equation}
In the Newtonian limit this reduces to
\begin{equation}
  H_{\rm ent}^{\rm Newt} = C' - \phi + {1 \over 2} (\Omega \times r)^2,
\end{equation}
where $\phi$ denotes the total gravitational field and $C'$ is an
integration constant which is the Newtonian limit of $C$.

For irrotational configurations Eq.~(\ref{eq:iEuler}) depends on the
solution of the continuity equation, which can only be solved in the
stellar interior.
In this case we cannot extend the definition of the enthalpy to
regions outside the neutron star, meaning that a straight-forward
definition of the effective potential is possible only for corotating
configurations.

\begin{figure*}[ht]
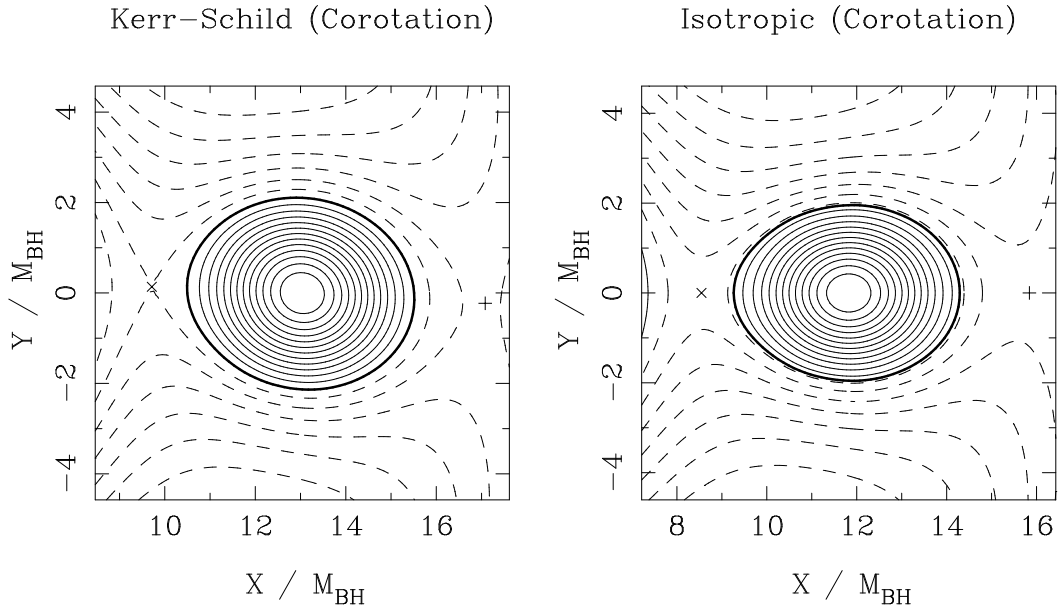

\vspace{0.5cm}
\begin{center}
  \includegraphics[height=8cm]{fig16a_ent_K005c.eps}
  \hspace{10pt}
  \includegraphics[height=8cm]{fig16b_ent_C005c.eps}
\end{center}
\caption{Contours of the enthalpy field $H_{\rm ent}$ extended
outside of the neutron star, which has a mass $\bar{M}_{\rm NS}=0.05$
and is corotating.
The thick solid curve denotes the stellar surface, thin
solid curves are contours located inside the neutron star, and dashed curves
those outside.
The left and right panels show configurations in the Kerr-Schild and
isotropic backgrounds, respectively.
Both depict the innermost point along the respective sequences,
listed in the last line of the corresponding tables.
The symbols $\times$ and $+$ denote the inner and outer
Lagrange points, respectively.
Note that the horizontal axis $X$ in the left panel is
the $X$ coordinate of the Kerr-Schild coordinate system
and that in the right panel is in the isotropic coordinate system.
}
\label{fig:ent_005c}
\end{figure*}

\begin{figure*}[ht]
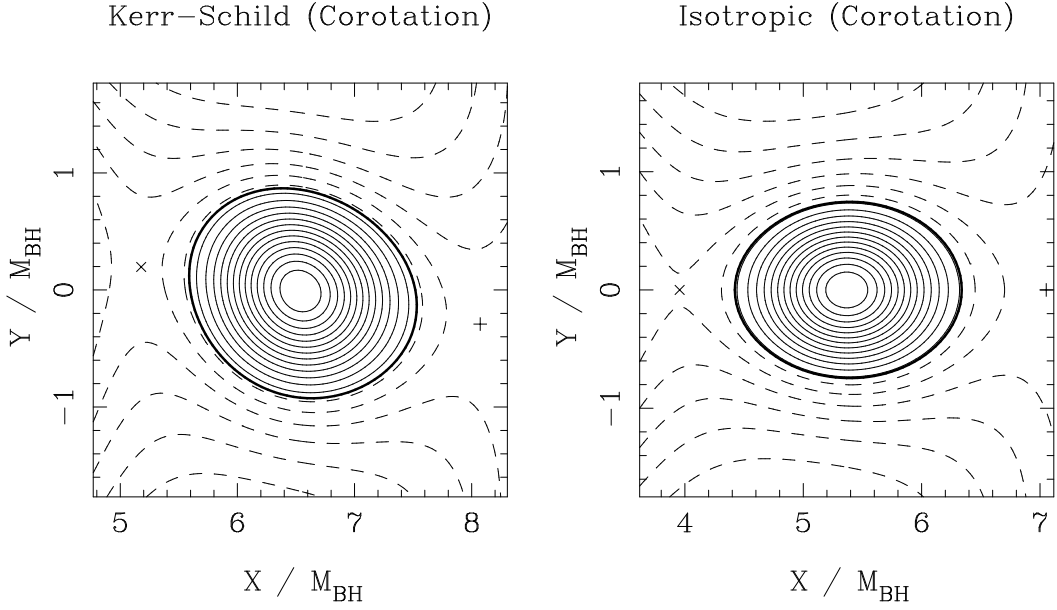

\vspace{0.5cm}
\begin{center}
  \includegraphics[height=8cm]{fig17a_ent_K01c.eps}
  \hspace{10pt}
  \includegraphics[height=8cm]{fig17b_ent_C01c.eps}
\end{center}
\caption{Same as Fig. \ref{fig:ent_005c}, but for a neutron star of
mass $\bar{M}_{\rm NS}=0.1$.}
\label{fig:ent_01c}
\end{figure*}

We show contours of the effective enthalpy field (\ref{eq:effect_ent})
in Figs. \ref{fig:ent_005c} and \ref{fig:ent_01c}
for neutron star of mass $\bar{M}_{\rm NS}=0.05$ and 0.1, respectively.
In each figure, the left panel is for the Kerr-Schild background
and the right panel for the isotropic one.
In both figures, we show the position of the inner Lagrange point using
the symbol ``$\times$'' and that of the outer one with ``$+$''.
The equipotential surface passing through the inner Lagrange point
defines the relativistic Roche lobe.
It is clear that the neutron star in our results still fits well
within its Roche lobe, but if we decrease the separation only a few
percent further, we expect a rapid deformation on the inner edge as a
cusp forms at the mass-shedding limit.

For the Kerr-Schild background the absence of a symmetry across the
$X\!\!-\!\!Z$ plane is again quite noticeable, in that the Lagrange points
do not lie on the $X$-axis.
For an isotropic background, on the other hand, the presence of this
symmetry force the Lagrange points to lie on the $X$-axis.

\subsection{Shift vector}

\begin{figure*}[ht]
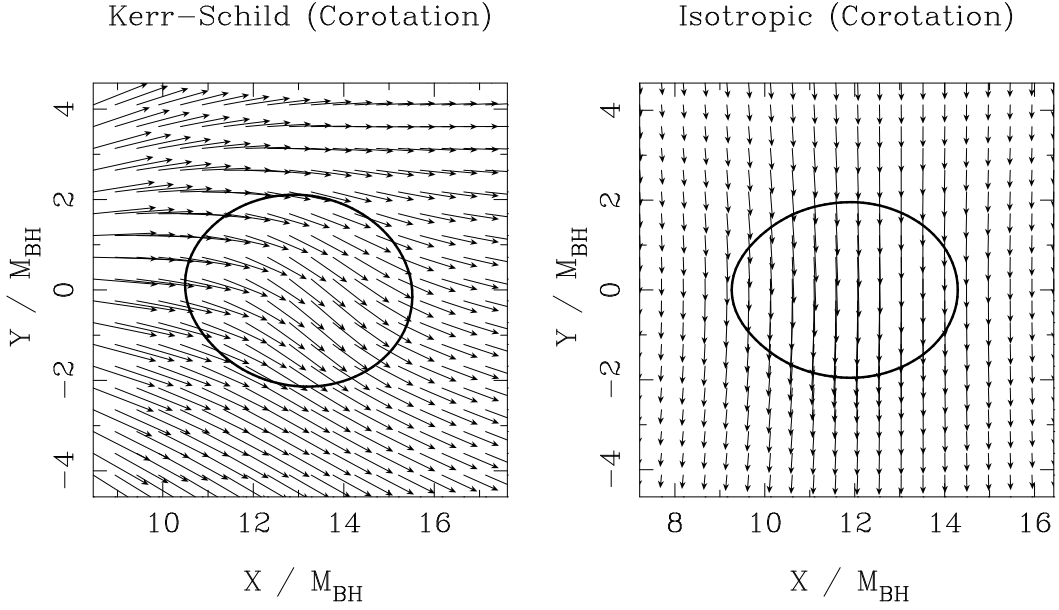

\vspace{0.5cm}
\begin{center}
  \includegraphics[height=8cm]{fig18a_shift_K005c.eps}
  \hspace{10pt}
  \includegraphics[height=8cm]{fig18b_shift_C005c.eps}
\end{center}
\caption{Shift vectors seen by an inertial observer
in the $X\!\!-\!\!Y$ plane (equatorial plane).  
The left and right panels show the Kerr-Schild and isotropic backgrounds,
respectively.
In both cases the neutron star has a mass $\bar{M}_{\rm NS}=0.05$,
and is corotating.
Thick solid circles in each panel denote the stellar surface,
and the arrows the direction of the shift vector.
Note that the horizontal axis $X$ in the left panel is
the $X$ coordinate of the Kerr-Schild coordinate system
and that in the right panel is in the isotropic coordinate system.}
\label{fig:shift_005c}
\end{figure*}

One of the most significant 
differences between the Kerr-Schild coordinate system
and the isotropic one is the existence/absence of the
black hole shift vector.
In isotropic coordinates the background shift is zero, while for
a Kerr-Schild background the main contribution to the total shift
arises from the Schwarzschild background.
We show the shift vector for an inertial observer,
$\beta^i_{\rm NS} +\beta^i_{\rm BH}$, for corotating
configurations with  $\bar{M}_{\rm NS}=0.05$ in Fig. \ref{fig:shift_005c}.
The left panel is for the Kerr-Schild case at the orbital separation
$X_{\rm NS}/M_{\rm BH}=13.0$ and the right panel for the isotropic case
at $X_{\rm NS}/M_{\rm BH}=11.8$.
The neutron star rotates counterclockwise.
It is evident that for the Kerr-Schild background the shift is dominated
by the outward-pointing background contribution $\beta^i_{\rm BH}$.
The lack of a symmetry across the $X\!\!-\!\!Z$ plane is again quite
obvious for the Kerr-Schild background.

\section{Summary}

We have computed quasiequilibrium sequences of BHNS binary systems
in general relativity.
Under the assumption of an extreme mass ratio, $M_{\rm BH} \gg M_{\rm NS}$,
we have treated the contribution of the black hole gravitational field
as a fixed background, adopting the Schwarzschild metric
in both Kerr-Schild and isotropic coordinates.
The neutron star is modeled as a relativistic polytropic EOS with adiabatic
index $\Gamma=2$.
We have solved a set of equations for two rotation states, irrotation and
corotation.
These results generalize those of BSS, which presented the first
relativistic, self-consistent method to evaluate quasiequilibrium
black hole-neutron star configurations.

We have developed a new numerical code for the present study, based on
a spectral methods numerical approach that was used for a series of works
on binary neutron stars
\cite{GourGTMB01,TanigGB01,TanigG02a,TanigG02b,TanigG03,BejgGGHTZ05}.
After confirming the validity of the numerical code in several tests,
we have computed quasiequilibrium sequences of the BHNS binaries with
a mass ratio $M_{\rm BH}/M_{\rm NS}=10$ and neutron star masses
$\bar{M}_{\rm NS}=0.05$, 0.1, and 0.15.
We expect that spectral methods allow us to construct each individual
model with greater accuracy than the finite difference methods used by BSS.
However, the assumption of smoothness inherent in spectral methods
prevents us from constructing models very close to the onset of tidal
disruption.
This leads to errors in parameters describing the onset of tidal
disruption that are greater than those obtained with finite differencing.
While we maintain here some of the assumptions from BSS, including extreme
mass ratios and polytropic equations of state,
and plan to relax these in the near future.

In agreement with earlier studies (e.g.~\cite{WiggL00}),
we find that the effect of rotation on the onset of tidal disruption is
fairly small and in the order of a few percent.
This does not rule out, however, that rotation has a larger effect on the
dynamics of the tidal disruption itself (compare \cite{FaberBSTR05}).
In particular, the rotational state of the neutron star may well
affect the size (or existence) of an accretion disk that may form as
the neutron star is disrupted.
Such an accretion disk is at the core of BHNS models as central engines
of short-period gamma ray bursts (e.g.~\cite{JankaERF99,Rossw05,CMiller05}).

We also find that the choice of background does have some effect on
coordinate-independent quantities describing the resulting binary
configurations.
This indicates that the choice of background may affect the degree to
which the solutions approximate quasiequilibrium,
as well as the amount of spurious gravitational radiation inherent
in the solutions.
It will be very interesting to study these differences in future
dynamical simulations (compare \cite{FaberBSTR05}).

\acknowledgments

It is a pleasure to thank Eric Gourgoulhon for useful comments on the
development of the numerical code.
TWB gratefully acknowledges support from
the J.~S.~Guggenheim Memorial Foundation.
JAF is supported by an NSF Astronomy and Astrophysics Postdoctoral
Fellowship under award AST-0401533.
This paper was supported in part by NSF Grants PHY-0205155 and 
PHY-0345151, and NASA Grant NNG04GK54G, to
University of Illinois at Urbana-Champaign,
and NSF Grant PHY 0139907 to Bowdoin College

\begin{widetext}
\appendix

\section{Equations} \label{app:eq_detail}

In this appendix, we present the explicit forms of equations
which we solve in our numerical code.
Having applied the decompositions of metric quantities
(\ref{eq:dec_nu}), (\ref{eq:dec_sigma}), and (\ref{eq:dec_shift})
to the equations (\ref{eq:logn}), (\ref{eq:lognp}), and
(\ref{eq:shift}),
we can derive the final forms for the neutron star components in a
numerically convenient form, inserting the black hole components as
a fixed background.

\subsection{Kerr-Schild background} \label{app:eq_ks}

\subsubsection{Gravitational field equations}

The equations (\ref{eq:logn}), (\ref{eq:lognp}), and (\ref{eq:shift}),
respectively, can be written as
\begin{eqnarray}
  &&\underline{\Delta} \nu_{\rm NS} = 4\pi \psi_{\rm NS}^4 (\rho +S)
  +2 \alpha_{\rm BH}^2 H_{\rm BH} l^i l^j
  \bar{\nabla}_i \bar{\nabla}_j \nu_{\rm NS} \nonumber \\
  &&\hspace{10pt}+{\alpha_{\rm BH}^4 H_{\rm BH} \over r_{\rm BH}}
  \Bigl[ 2(1+4H_{\rm BH}) l^i \bar{\nabla}_i \nu_{\rm NS} -l^i \bar{\nabla}_i
    \sigma_{\rm NS} \Bigr] \nonumber \\
  &&\hspace{10pt}+{\psi_{\rm NS}^{-8} \over \alpha_{\rm BH}^2}
  \tilde{A}_{ij}^{\rm NS} \tilde{A}_{\rm NS}^{ij}
  -{4 \psi_{\rm NS}^{-2} \alpha_{\rm BH}^2 H_{\rm BH} \over
    3 \alpha_{\rm NS} r_{\rm BH}} (2+3H_{\rm BH}) (3+4H_{\rm BH}) l_i l_j
  \tilde{A}_{\rm NS}^{ij} \nonumber \\
  &&\hspace{10pt}-(\eta^{ij} -2\alpha_{\rm BH}^2 H_{\rm BH}l^i l^j)
  (\bar{\nabla}_i \nu_{\rm NS}) (\bar{\nabla}_j \sigma_{\rm NS})
  -{2 \psi_{\rm NS}^4 \alpha_{\rm BH}^4 H_{\rm BH} \over \alpha_{\rm NS}
    r_{\rm BH}^2} (2+10H_{\rm BH}+9H_{\rm BH}^2) l_i \beta^i_{\rm NS}
  \nonumber \\
  &&\hspace{10pt}+{4 \alpha_{\rm BH}^6 H_{\rm BH}^2 \over 3r_{\rm BH}^2}
  \Bigl[ 2 \Bigl( {\psi_{\rm NS}^4 \over \alpha_{\rm NS}^2} -1 \Bigr)
    (2+3H_{\rm BH})^2 +(\psi_{\rm NS}^4 -1) (1+3H_{\rm BH})^2 \nonumber \\
  &&\hspace{70pt}-3 \Bigl( {\psi_{\rm NS}^4 \over \alpha_{\rm NS}} -1 \Bigr)
    (2+10H_{\rm BH}+9H_{\rm BH}^2) \Bigr], \label{eq:logn_ks}
\end{eqnarray}
\begin{eqnarray}
  &&\underline{\Delta} \sigma_{\rm NS} = 4\pi \psi_{\rm NS}^4 S
  +2 \alpha_{\rm BH}^2 H_{\rm BH} l^i l^j
  \bar{\nabla}_i \bar{\nabla}_j \sigma_{\rm NS} +{\alpha_{\rm BH}^4 H_{\rm BH}
    \over r_{\rm BH}}
  \Bigl[ 2(1+4H_{\rm BH}) l^i \bar{\nabla}_i \sigma_{\rm NS}
    -l^i \bar{\nabla}_i \nu_{\rm NS} \Bigr] \nonumber \\
  &&\hspace{10pt}+{3\psi_{\rm NS}^{-8} \over 4\alpha_{\rm BH}^2}
  \tilde{A}_{ij}^{\rm NS} \tilde{A}_{\rm NS}^{ij}
  -{\psi_{\rm NS}^{-2} \alpha_{\rm BH}^2 H_{\rm BH} \over
    \alpha_{\rm NS} r_{\rm BH}} (2+3H_{\rm BH}) (3+4H_{\rm BH}) l_i l_j
  \tilde{A}_{\rm NS}^{ij} \nonumber \\
  &&\hspace{10pt}-{1 \over 2} (\eta^{ij}
  -2\alpha_{\rm BH}^2 H_{\rm BH} l^i l^j) \Bigl[
    (\bar{\nabla}_i \nu_{\rm NS}) (\bar{\nabla}_j \nu_{\rm NS})
    +(\bar{\nabla}_i \sigma_{\rm NS}) (\bar{\nabla}_j \sigma_{\rm NS}) \Bigr]
  \nonumber \\
  &&\hspace{10pt}-{2 \psi_{\rm NS}^4 \alpha_{\rm BH}^4 H_{\rm BH} \over
    \alpha_{\rm NS}
    r_{\rm BH}^2} (2+10H_{\rm BH}+9H_{\rm BH}^2) l_i \beta^i_{\rm NS}
  \nonumber \\
  &&\hspace{10pt}+{2 \alpha_{\rm BH}^6 H_{\rm BH}^2 \over r_{\rm BH}^2}
  \Bigl[ \Bigl( {\psi_{\rm NS}^4 \over \alpha_{\rm NS}^2} -1 \Bigr)
    (2+3H_{\rm BH})^2 +(\psi_{\rm NS}^4 -1) (1+3H_{\rm BH})^2 \nonumber \\
  &&\hspace{70pt}-2 \Bigl( {\psi_{\rm NS}^4 \over \alpha_{\rm NS}} -1 \Bigr)
    (2+10H_{\rm BH}+9H_{\rm BH}^2) \Bigr], \label{eq:lognp_ks}
\end{eqnarray}
\begin{eqnarray}
  &&\underline{\Delta} \beta^i_{\rm NS} +{1 \over 3} \bar{\nabla}^i
  \bar{\nabla}_j \beta^j_{\rm NS}
  =16\pi \alpha_{\rm NS} \alpha_{\rm BH} \psi_{\rm NS}^4 j^i
  +{2 \over 3} \alpha_{\rm BH}^2 H_{\rm BH} \Bigl( 3 l^j l^k \bar{\nabla}_j
  \bar{\nabla}_k \beta^i_{\rm NS} +l^i l^k \bar{\nabla}_k \bar{\nabla}_j
  \beta^j_{\rm NS} \Bigr) \nonumber \\
  &&\hspace{10pt}-{\alpha_{\rm BH}^2 H_{\rm BH} \over r_{\rm BH}}
  \Bigl[ 4l^i \bar{\nabla}_j \beta^j_{\rm NS} -\alpha_{\rm BH}^2
    (3+8H_{\rm BH}) l^j \bar{\nabla}_j \beta^i_{\rm NS} \nonumber \\
  &&\hspace{70pt}-{1 \over 3} \Bigl\{ \eta^{ij} +2\alpha_{\rm BH}^2
    (9+11H_{\rm BH})
    l^i l^j \Bigr\} l_k \bar{\nabla}_j \beta^k_{\rm NS} \Bigr] \nonumber \\
  &&\hspace{10pt}-{2\alpha_{\rm BH}^4 H_{\rm BH} \over 3 r_{\rm BH}^2}
  \Bigl[ (4+11H_{\rm BH}) \beta^i_{\rm NS} -\alpha_{\rm BH}^2
    (12 +51 H_{\rm BH} +46 H_{\rm BH}^2) l^i l_j \beta^j_{\rm NS} \Bigr]
  \nonumber \\
  &&\hspace{10pt}-{8 \alpha_{\rm BH}^8 H_{\rm BH} \over 3 r_{\rm BH}^2}
  (\alpha_{\rm NS} -1) (2+10H_{\rm BH}+9H_{\rm BH}^2) l^i
  -{2 \alpha_{\rm NS} \over \psi_{\rm NS}^6}
  \tilde{A}_{\rm NS}^{ij} \bar{\nabla}_j (3\sigma_{\rm NS} -4\nu_{\rm NS})
  \nonumber \\
  &&\hspace{10pt}+{2 \alpha_{\rm NS} \alpha_{\rm BH}^2 H_{\rm BH} \over
    \psi_{\rm NS}^6
    r_{\rm BH}} \tilde{A}_{\rm NS}^{ij} l_j \nonumber \\
  &&\hspace{10pt}-{4 \alpha_{\rm BH}^6 H_{\rm BH} \over 3 r_{\rm BH}}
  (2+3H_{\rm BH}) \Bigl[ (1+2H_{\rm BH}) \eta^{ij}
    -(3+2H_{\rm BH}) l^i l^j \Bigr]
  \bar{\nabla}_j (3\sigma_{\rm NS} -4 \nu_{\rm NS}), \label{eq:shift_ks}
\end{eqnarray}
where $\underline{\Delta}$ and $\bar{\nabla}_i$ are the flat Laplace
operator and partial derivative.
The conformally related trace-free extrinsic curvature of the neutron star
part is defined as
\begin{equation}
  \tilde{A}_{\rm NS}^{ij} \equiv {\psi_{\rm NS}^6 \over 2 \alpha_{\rm NS}}
  \Bigl( \tilde{D}^i \beta^j_{\rm NS} +\tilde{D}^j \beta^i_{\rm NS}
  -{2 \over 3} \tilde{\gamma}^{ij} \tilde{D}_k \beta^k_{\rm NS} \Bigr).
\end{equation}

\subsubsection{Equation of continuity} \label{app:eoc_ks}

Having defined a velocity field related to the orbital motion, i.e.,
\begin{equation}
  W^i \equiv \psi^4 h \gamma_{\rm n} U_0^i, \label{eq:wi}
\end{equation}
we can express the velocity potential as
\begin{equation}
  \Psi = \Psi_0 +\eta_{ij} W_0^i x^j,
\end{equation}
where $W_0^i$ is the constant value of $W^i$ at the center of
the neutron star, i.e.,
\begin{equation}
  W_0^i \equiv ( \psi^4 h \gamma_{\rm n} U_0^i )_{\rm center}
  = {\rm constant}. \label{eq:wi0}
\end{equation}
Since the Newtonian limit of the right hand side of 
Eq. (\ref{eq:wi}) is the orbital motion
$(\Omega \times r_{\rm BH})^i$, Eq. (\ref{eq:wi0}) is merely the relativistic
analogue of the translational motion of the center of the neutron star.
Thus, the quantity $\Psi_0$ is regarded as the residual of the velocity
potential, once the constant rotational component is subtracted away.
The gradient of $\Psi_0$ yields the counter-rotation 
seen by a co-orbiting observer.
The gradient and the Laplacian of $\Psi_0$ become
\begin{eqnarray}
  \bar{\nabla}^i \Psi &=& \bar{\nabla}^i \Psi_0 + W_0^i, \\
  \underline{\Delta} \Psi &=& \underline{\Delta} \Psi_0.
\end{eqnarray}
Inserting the Kerr-Schild metric for the black hole into Eq.~(\ref{eq:eoc_t}),
we have our final form of the equation of continuity,
\begin{eqnarray}
  &&\zeta H_{\rm ent} \underline{\Delta} \Psi_0
  +\Bigl[ (1-\zeta H_{\rm ent}) (\bar{\nabla}^i H_{\rm ent}
    -2\alpha_{\rm BH}^2 H_{\rm BH} l^i l^j \bar{\nabla}_j H_{\rm ent})
  \nonumber \\
  &&\hspace{70pt}+\zeta H_{\rm ent} (\bar{\nabla}^i \sigma_{\rm NS}
    -2 \alpha_{\rm BH}^2 H_{\rm BH} l^i l^j \bar{\nabla}_j \sigma_{\rm NS})
    \nonumber \\
  &&\hspace{70pt}-2 \zeta H_{\rm ent} {\alpha_{\rm BH}^4 H_{\rm BH} \over
      r_{\rm BH}} (1 +4 H_{\rm BH}) l^i \Bigr] \bar{\nabla}_i \Psi_0
  \nonumber \\
  &&= 2 \zeta H_{\rm ent} \alpha_{\rm BH}^2 H_{\rm BH} l^i l^j
  \bar{\nabla}_i \bar{\nabla}_j \Psi_0
  +(W^i -W_0^i) \bar{\nabla}_i H_{\rm ent}
  +\zeta H_{\rm ent} {W^i \over \gamma_{\rm n}} \bar{\nabla}_i
  \gamma_{\rm n} \nonumber \\
  &&\hspace{10pt}+2\zeta H_{\rm ent} \psi^4_{\rm NS} h \gamma_{\rm n}
	{\alpha_{\rm BH}^3 H_{\rm BH} \over r_{\rm BH}} (1 +3H_{\rm BH})
	\nonumber \\
  &&\hspace{10pt}+\Bigl[ \zeta H_{\rm ent} \bar{\nabla}_i
    (H_{\rm ent} -\sigma_{\rm NS}) -2\alpha_{\rm BH}^2 H_{\rm BH} l_i
    \Bigl\{ \zeta H_{\rm ent} l^j \bar{\nabla}_j
    (H_{\rm ent} -\sigma_{\rm NS}) -l^j \bar{\nabla}_j
    H_{\rm ent} \Bigr\} \nonumber \\
  &&\hspace{30pt}+2 \zeta H_{\rm ent} {\alpha_{\rm BH}^4 H_{\rm BH} \over
      r_{\rm BH}} (1 +4H_{\rm BH}) l_i \Bigr] W_0^i.
  \label{eq:conti_ks}
\end{eqnarray}

\subsubsection{Determination of the orbital angular velocity}
\label{app:ome_ks}

The left-hand side of the equation (\ref{eq:orbit}) can be expressed 
\begin{eqnarray}
  {\partial \over \partial X} \ln \gamma_0 \Bigl|_{(X_{\rm NS},0,0)}
  &=&{1 \over 2} \Bigl[ 1 -{\psi^4 \over \alpha^2} \Bigl\{ (\beta^X_{\rm I})^2
    (1+2H_{\rm BH}) +(\beta^Y_{\rm I} +\Omega X_{\rm NS})^2
    +(\beta^Z_{\rm I})^2 \Bigr\} \Bigr]^{-1} \nonumber \\
  &&\times \Bigl[ {\partial \over \partial X} \Bigl( {\psi^4 \over \alpha^2}
    \Bigr) \Bigl\{ (\beta^X_{\rm I})^2 (1+2H_{\rm BH})
    +(\beta^Y_{\rm I} +\Omega X_{\rm NS})^2 +(\beta^Z_{\rm I})^2
    \Bigr\} \nonumber \\
  &&\hspace{10pt}+2 \Bigl( {\psi^4 \over \alpha^2} \Bigr) \Bigl\{
    \beta^X_{\rm I} {\partial \beta^X_{\rm I} \over \partial X}
    (1+2H_{\rm BH}) -(\beta^X_{\rm I})^2 {H_{\rm BH} \over r_{\rm BH}}
    \nonumber \\
  &&\hspace{60pt}+(\beta^Y_{\rm I} +\Omega X_{\rm NS})
    \Bigl( {\partial \beta^Y_{\rm I} \over \partial X} +\Omega \Bigr)
    +\beta^Z_{\rm I} {\partial \beta^Z_{\rm I} \over \partial X} \Bigr\}
    \Bigr] \Bigl|_{(X_{\rm NS},0,0)},
\end{eqnarray}
where $\beta^i_{\rm I}$ denotes the shift vector seen by the inertial
observer, i.e., $\beta^i_{\rm I} \equiv \beta^i_{\rm NS} +\beta^i_{\rm BH}$.

The orbital angular velocity appears explicitly
in the above equation, which corresponds to
the left-hand side of Eq. (\ref{eq:orbit}).
On the contrary, we compute the right-hand side of Eq. (\ref{eq:orbit})
in the form shown in which it is shown.

\subsection{Isotropic background} \label{app:eq_cf}

\subsubsection{Gravitational field equations}

In the case of isotropic backgrounds,
$\tilde{R}_{ij}=\tilde{R}=\tilde{A}^{ij}_{\rm BH}=K=0$,
because $\beta^i_{\rm BH}=0$ and $\tilde{\gamma}_{ij}=\eta_{ij}$.
This reduces the field equations to a much more simple form than those
in Kerr-Schild backgrounds.

The equations (\ref{eq:logn}), (\ref{eq:lognp}), and (\ref{eq:shift})
are re-written as follows.
\begin{eqnarray}
  \underline{\Delta} \nu_{\rm NS} &=&4 \pi \psi^4 (\rho +S)
  +\psi^{-8} \tilde{A}_{ij}^{\rm NS} \tilde{A}^{ij}_{\rm NS}
  -(\bar{\nabla}_i \nu_{\rm NS}) (\bar{\nabla}^i \sigma_{\rm NS})
  \nonumber \\
  &&-{1 \over \displaystyle \Bigl( 1-
    {M_{\rm BH}^2 \over 4r_{\rm BH}^2} \Bigr)}
  {M_{\rm BH} \over r_{\rm BH}^3} X^i
  \Bigl[ {M_{\rm BH} \over 2r_{\rm BH}} (\bar{\nabla}_i \nu_{\rm NS})
    + (\bar{\nabla}_i \sigma_{\rm NS}) \Bigr],\label{eq:logn_cf}
\end{eqnarray}
\begin{eqnarray}
  \underline{\Delta} \sigma_{\rm NS} &=&4 \pi \psi^4 S
  +{3 \over 4} \psi^{-8} \tilde{A}_{ij}^{\rm NS} \tilde{A}_{\rm NS}^{ij}
  -{1 \over 2} \Bigl[ (\bar{\nabla}_i \nu_{\rm NS}) (\bar{\nabla}^i
    \nu_{\rm NS}) +(\bar{\nabla}_i \sigma_{\rm NS}) (\bar{\nabla}^i
    \sigma_{\rm NS}) \Bigr] \nonumber \\
  &&-{1 \over \displaystyle
    \Bigl( 1- {M_{\rm BH}^2 \over 4r_{\rm BH}^2} \Bigr)}
    {M_{\rm BH} \over r_{\rm BH}^3} X^i
    \Bigl[ (\bar{\nabla}_i \nu_{\rm NS})
      + {M_{\rm BH} \over 2r_{\rm BH}} (\bar{\nabla}_i \sigma_{\rm NS})
      \Bigr], \label{eq:lognp_cf}
\end{eqnarray}
\begin{equation}
  \underline{\Delta} \beta^i_{\rm NS} +{1 \over 3} \bar{\nabla}^i
  \bar{\nabla}_j
  \beta^j_{\rm NS} =16 \pi \alpha \psi^4 j^i -{2 \alpha \over \psi^6}
  \tilde{A}^{ij}_{\rm NS} \bar{\nabla}_j \Bigl[ 3(\sigma_{\rm NS}
    +\sigma_{\rm BH}) -4(\nu_{\rm NS} +\nu_{\rm BH}) \Bigr].
  \label{eq:shift_cf}
\end{equation}

\subsubsection{Equation of continuity} \label{app:eoc_cf}

The equation of continuity (\ref{eq:eoc_t}) is written
\begin{eqnarray}
  &&\zeta H_{\rm ent} \underline{\Delta} \Psi_0 + \Biggl[
    (1-\zeta H_{\rm ent}) \bar{\nabla}^i H_{\rm ent} +\zeta H_{\rm ent}
    \Biggl\{ \bar{\nabla}^i \sigma_{\rm NS} +{1 \over \displaystyle
      \Bigl( 1 -{M_{\rm BH}^2 \over 4r_{\rm BH}^2} \Bigr)}
	     {M_{\rm BH}^2 \over 2r_{\rm BH}^4}
    X^i \Biggr\} \Biggr] \bar{\nabla}_i \Psi_0 \nonumber \\
  &&\hspace{10pt}=(W^i -W^i_0) \bar{\nabla}_i H_{\rm ent} +\zeta H_{\rm ent}
  \Biggl[ W^i_0 \Biggl\{ \bar{\nabla}_i (H_{\rm ent} -\sigma_{\rm NS})
    -{1 \over \displaystyle \Bigl( 1 -{M_{\rm BH}^2 \over 4r_{\rm BH}^2}
      \Bigr) }
    {M_{\rm BH}^2 \over 2r_{\rm BH}^4} X_i \Biggr\} +{W^i \over \gamma_{\rm n}}
    \bar{\nabla}_i \gamma_{\rm n} \Biggr]. \nonumber \\
  \label{eq:conti_cf}
\end{eqnarray}
Here we have used the same quantities $\Phi_0$, $W^i$, and $W^i_0$
as defined in the Kerr-Schild part, Sec. \ref{app:eoc_ks}.

\subsubsection{Determination of the orbital angular velocity}
\label{app:ome_cf}

\begin{eqnarray}
  {\partial \over \partial X} \ln \gamma_0 \Bigl|_{(X_{\rm NS},0,0)}
  &=& {1 \over 2} \Bigl[ 1 -{\psi^4 \over \alpha^2} \Bigl\{
    (\beta^X_{\rm NS})^2 +(\beta^Y_{\rm NS} +\Omega X_{\rm NS})^2
    +(\beta^Z_{\rm NS})^2 \Bigr\} \Bigr]^{-1} \nonumber \\
  &&\times \Bigl[ {\partial \over \partial X} \Bigl(
    {\psi^4 \over \alpha^2} \Bigr) \Bigl\{ (\beta^X_{\rm NS})^2
    +(\beta^Y_{\rm NS} + \Omega X_{\rm NS})^2 +(\beta^Z_{\rm NS})^2
    \Bigr\} \nonumber \\
  &&\hspace{10pt} +\Bigl( {\psi^4 \over \alpha^2} \Bigr) \Bigl\{
      {\partial \over \partial X} \Bigl( (\beta^X_{\rm NS})^2
      +(\beta^Y_{\rm NS})^2 +(\beta^Z_{\rm NS})^2 \Bigr) \nonumber \\
  &&\hspace{50pt}+2 \Omega \Bigl( \beta^Y_{\rm NS}
      +X_{\rm NS} {\partial \beta^Y_{\rm NS}
	\over \partial X} \Bigr) +2 \Omega^2 X_{\rm NS} \Bigr\} \Bigr]
  \Bigl|_{(X_{\rm NS},0,0)}. \nonumber \\
\end{eqnarray}
Note here that since there is no black hole shift vector in the isotropic
background, we write the shift vector seen by the inertial observer as
$\beta^i_{\rm NS}$.
\end{widetext}



\begin{thebibliography}{99}
\bibitem{footnote}
We point out that we treat the Kerr-Schild and isotropic backgrounds
slightly differently.
For a Kerr-Schild background, the conformally related background
metric $\tilde \gamma_{ij}$ is identified with the spatial metric
$\gamma_{ij}$ of a Kerr-Schild metric.
For Schwarzschild in isotropic coordinates, the spatial metric is
$\gamma_{ij} = \psi^4_{\rm BH} \eta_{ij}$.
In this case we absorb $\psi_{\rm BH}$ in our conformal factor $\psi$
(see equation (\ref{eq:dec_conf})), and identify the conformally related
metric $\tilde \gamma_{ij}$ with the flat metric $\eta_{ij}$.
\end{thebibliography}
\end{document}